\documentclass{aa}  %este es la version comun a dos cols.

\usepackage{graphicx}
\usepackage{longtable}
\usepackage{txfonts}
\begin{document}

   \title{HD 80606: Searching the chemical signature of planet formation\thanks{The data presented
   herein were obtained at the W.M. Keck Observatory, which is operated as a scientific partnership
   among the California Institute of Technology, the University of California and the
   National Aeronautics and Space Administration. The Observatory was made possible by the generous
   financial support of the W.M. Keck Foundation.}
   \footnote{Reduced spectra of HD 80606 and HD 80607 (FITS files) are only available in electronic
   form at the CDS via anonymous ftp to cdsarc.u-strasbg.fr (130.79.128.5) or
   via http://cdsweb.u-strasbg.fr/cgi-bin/qcat?J/A+A/}
   }

   %\subtitle{I. Overviewing the $\kappa$-mechanism}
   
   \titlerunning{Searching the signature of planet formation}
   \authorrunning{Saffe et al.}

   \author{C. Saffe\inst{1,2}, M. Flores\inst{1} \and A. Buccino\inst{3,4}%\fnmsep\thanks{Just to show the usage of the elements in the author field}
          }

   \institute{Instituto de Ciencias Astron\'omicas, de la Tierra y del Espacio (ICATE-CONICET),
              C.C 467, 5400, San Juan, Argentina.
             \email{csaffe,mflores@icate-conicet.gob.ar}
         \and
         Universidad Nacional de San Juan (UNSJ), Facultad de Ciencias Exactas, F\'isicas y Naturales (FCEFN), San Juan, Argentina.
         \and
         Instituto de Astronom\'ia y F\'isica del Espacio (IAFE-CONICET), Buenos Aires, Argentina.
             \email{abuccino@iafe.uba.ar}
        \and
        Departamento de F\'isica, Facultad de Ciencias Exactas y Naturales (FCEN), Universidad de Buenos Aires (UBA), Buenos Aires, Argentina. 
        }

   \date{Received xxx, xxx ; accepted xxxx, xxxx}

% \abstract{}{}{}{}{} 
% 5 {} token are mandatory
 
  \abstract
  % context heading (optional)
  % {} leave it empty if necessary  
   {Binary systems with similar components are ideal laboratories which allow 
   several physical processes to be tested, such as the possible chemical pattern
   imprinted by the planet formation process.}
  % aims heading (mandatory)
   {We explore the probable chemical signature of planet formation in the remarkable binary
   system {HD 80606 - HD 80607}. The star {HD 80606} hosts a giant planet with {$\sim$4 M$_{Jup}$}
   detected by both transit and radial velocity techniques, being one of the most eccentric
   planets detected to date. We study condensation temperature T$_{c}$ trends of volatile
   and refractory element abundances to determine whether there is a depletion
   of refractories that could be related to the terrestrial planet formation.}
  % methods heading (mandatory)
   {We carried out a high-precision abundance determination in both components of the binary system,
   using a line-by-line strictly differential approach, using the Sun as a reference and then
   using {HD 80606} as reference. The stellar parameters T$_{eff}$, {log g}, [Fe/H] and v$_{turb}$
   were determined by imposing differential ionization and excitation equilibrium of Fe I and Fe II lines,
   using an updated version of the program FUNDPAR, together with 1D LTE ATLAS9 model atmospheres and the MOOG code.
   Then, we derived detailed abundances of 24 different species using equivalent widths and spectral
   synthesis with the program MOOG. The chemical patterns were compared with the solar-twins
   T$_{c}$ trends of \citet{melendez09} and
   with a sample of solar-analog stars with {[Fe/H]$\sim$+0.2 dex} from \citet{neves09}.
   The T$_{c}$ trends were also compared mutually between both stars of the binary system.}
  % conclusions heading (optional), leave it empty if necessary 
   {From the study of T$_{c}$ trends, we concluded that the stars {HD 80606} and {HD 80607} do not
   seem to be depleted in refractory elements, which is different for the case of the Sun.
   Then, following the interpretation of \citet{melendez09}, the terrestrial planet formation
   would have been less efficient in the components of this binary system than in the Sun.
   The lack of a trend for refractory elements with T$_{c}$ between both stars implies that
   the presence of a giant planet do not neccesarily imprint a chemical signature in their host stars,
   similar to the recent result of \citet{liu14}. This is also in agreement with \citet{melendez09},
   who suggest that the presence of close-in giant planets might prevent the formation of terrestrial planets.
   Finally, we speculate about a possible (ejected or non-detected) planet around the star {HD 80607}.}
   {}
   

   \maketitle
%
%________________________________________________________________

\section{Introduction}

%The presence of low-mass companions or exoplanets may imprint chemical signatures
%in the photospheres of their host stars.
Main-sequence stars with giant planets are, on average, metal-rich
compared to stars without planetary mass companions \citep[e.g. ][]{santos04,santos05,fischer-valenti05}.
On the other hand, Neptune-like or super-Earth planets do not
seem to be formed preferentially around metal-rich stars \citep[e.g. ][]{udry06,sousa08}.
\citet[][ hereafter M09]{melendez09} have further suggested that small chemical anomalies (rather than
a global excess of metallicity) are a possible signature of terrestrial planet formation.
The authors showed that the Sun is deficient in refractory elements relative to volatile
when compared to solar twins, suggesting that the refractory elements depleted in the
solar photosphere are possibly locked up in terrestrial planets and/or in the cores of
giant planets.

Most binary stars are believed to have formed from a common molecular cloud.
This is supported both by observations of binaries in star forming regions
\citep[e.g.][]{reipurth07,vogt12,king12} and by numerical models of binary formation
\citep[e.g.][]{reipurth-mikkola12,kratter11}.
These systems are ideal laboratories to look for possible chemical differences
between their components, specially for physically similar stars which help to minimize the errors.
For the case of main-sequence stars, \citet{desidera04} studied the components of
23 wide binary stars and showed that
most pairs present almost identical abundances, with only 4 pairs showing differences between
0.02 dex and 0.07 dex. A similar conclusion was reached by \citet{desidera06}, showing that
only 6 out of 33 southern binary stars with similar components present differences between
0.05 and 0.09 dex.
The origin of the slight differences in these few cases is not totally clear,
and a possible explanation is the planet formation process \citep[e.g. ][]{gratton01,desidera04,desidera06}.

There are very few binary systems with similar components (where one of them host a planet)
studied in the literature, comparing in detail the chemical composition between them.
For instance, the binary system {16 Cyg} is composed of a pair of stars with spectral types
{G1 V + G2 V}, and the B component hosts a giant planet of {$\sim$1.5 M$_{Jup}$} \citep{cochran97}.
This system have received the attention of many different abundance works.
\citet{takeda05} and \citet{schuler11}
suggested that both stars present the same chemical composition,
while other studies found that {16 Cyg A} is more metal-rich than the B component 
\citep{laws-gonzalez01,ramirez11,tuccimaia14}.
In particular, \citet{tuccimaia14} also find a trend between refractories and the
condensation temperature T$_{c}$, which could be interpreted as a signature of the rocky
accretion core of the giant planet {16 Cyg Bb}.
Another example is the binary system {HAT-P-1} composed of an {F8 V + G0 V} pair,
in which the cooler star hosts a {$\sim$0.53 M$_{Jup}$} transiting planet \citep{bakos07}.
Recently, \citet{liu14} found almost the same chemical abundances on both stars and concluded
that the presence of giant planets does not necessarily imply differences in their composition.
Both members of the binary system present an identical positive correlation with T$_{c}$, suggesting
that the terrestrial formation process was probably less efficient in this system.
\citet{liu14} also discuss why the chemical signature of planet formation is detected in 
the binary system {16 Cyg} but not in the {HAT-P-1} system. 
The planet {16 Cyg Bb} {($\sim$1.5 M$_{Jup}$)} is more massive than
the planet {HAT-P-1 Bb} {($\sim$0.5 M$_{Jup}$)}, allowing to imprint the chemical signature
in their host stars. The stellar masses in the binary system {HAT-P-1}
\citep[1.16 and 1.12 M$_{\sun}$,][]{bakos07} are slightly higher than in the system {16 Cyg} 
\citep[1.05 and 1.00 M$_{\sun}$,][]{ramirez11}.
This implies less massive convection zones in the stars of the system {HAT-P-1} 
(i.e. more prone to imprint the chemical signature) but also shorter pre-main-sequence disc lifetimes
(i.e. more difficult to imprint the chemical signature). These points illustrate how complicated and
challenging could be to determine the possible effects of planet formation using stellar abundances.
Then, there is a need for additional stars hosting planets in binary systems to
be compared through a high-precision abundance determination.

Using radial-velocity measurements, \citet{naef01} detected first a giant planet around
the solar-type star {HD 80606}, which is the primary of the wide binary system {HD 80606 - HD 80607}
(components A and B). To date, there is no planet detected around
the B component. The separation between A and B stars is 21.1" \citep[e.g. ][]{dn02},
corresponding to $\sim$1000 AU at the distance of about 60 pc \citep{laughlin09}.
This binary system is particularly notable for several reasons.
Both stars present very similar fundamental parameters (their effective temperatures differ
only in {67 K} and their superficial gravities in {0.01 dex}, as we see later).
The reported spectral types are {G5 V + G5 V}, as described in the Hipparcos catalog.
This makes this system a new member of the selected group of binaries with very similar components.
The exoplanet {HD 80606 b} have a period of 111.8 days and one of the most eccentric orbits to date \citep[{e = 0.927},][]{naef01},
probably due to the influence of the B star \citep{wu-murray03}.
Besides the radial-velocity detection, \citet{laughlin09} reported a secondary transit
for {HD 80606 b} using 8 $\mu$m {\it{Spitzer}} observations, while \citet{moutou09}
detected the primary transit of the planet and measured a planet radius of 0.9 M$_{Jup}$.
Then, future observations of the atmosphere of this transiting planet could be compared to
the natal chemical environment established by a binary star elemental abundances, as suggested
by \citet{teske13}.
These significant features motivated this study, exploring the possible chemical signature
of planet formation in this remarkable system.

There are some previous abundance measurements of {HD 80606} in the literature.
A number of elements show noticeable discrepancies in the reported values.
Notably, using the same stellar parameters, the Na abundance have been reported as
{+0.30$\pm$0.05 dex} and {+0.53$\pm$0.12 dex} \citep{beirao05,mortier13} while the Si abundance
resulted {+0.40$\pm$0.09 dex} and {+0.27$\pm$0.06 dex} \citep{mortier13,gilli06}.
These differences also encouraged this work. We perform a high-precision abundance study
analyzing both members of this unique binary system using a line-by-line differential approach,
aiming to detect a slight contrast between their components.

This work is organized as follows.
In Section 2 we describe the observations and data reduction, while in Section 3 we present
the stellar parameters and chemical abundance analysis. In Section 4 we show the results and
discussion, and finally in Section 5 we highlight our main conclusions.

\section{Observations and data reduction}

Stellar spectra of {HD 80606} and {HD 80607} were obtained with the
High Resolution Echelle Spectrometer (HIRES) attached on the right Nasmyth
platform of the Keck 10-meter telescope on Mauna Kea, Hawaii. 
The slit used was B2 with a width of 0.574 arcsec, which provides a measured
resolution of $\sim$67000 at $\sim$5200 \AA\footnote{http://www2.keck.hawaii.edu/inst/hires/slitres.html}.
The spectra were downloaded from the Keck Observatory
Archive (KOA)\footnote{http://www2.keck.hawaii.edu/koa/koa.html}, under the program ID A271Hr.

The observations were taken on March, 15th 2011 with {HD 80607} observed
immediately after {HD 80606}, using the same spectrograph configuration.
The exposure times were 3 x 300 s for both targets.
We measured a S/N $\sim$ 330 for each of the binary components.
The asteroid Iris was also observed with the same spectrograph setup
achieving a similar S/N, to acquire the solar spectrum useful for
reference in our (initial) differential analysis. We note however that
the final differential study with the highest abundance precision is between
{HD 80606} and {HD 80607}, due to their high degree of similarity.

Our resolving power is $\sim\%$40 higher than those reported in previous works
\citep{ecuvillon06,gilli06,mortier13}.
However, even for a similar resolution and S/N, the differential
line-by-line approach applied here results in a significant improvement
on the derived abundances, as we show in the next sections.

We reduced the HIRES spectra using the data reduction package
MAKEE\footnote{http://www.astro.caltech.edu/~tb/makee/}
(MAuna Kea Echelle Extraction), which performs the usual reduction process
including bias subtraction, flat fielding, spectral order extractions, and
wavelength calibration. The continuum normalization and other
operations (Doppler correction and combining spectra) was perfomed using
IRAF\footnote{IRAF is distributed by the National Optical Astronomical
Observatories which is operated by the Association of Universities for Research
in Astronomy, Inc., under a cooperative agreement with the National Science Foundation.}.

\section{Stellar parameters and chemical abundance analysis}

We start by measuring the equivalent widths (EW) of Fe I and Fe II lines in the spectra
of our program stars using the IRAF task {\it{splot}}, and then continued with other
chemical species. The lines list and relevant laboratory data (such as excitation potential and
oscilator strengths) were taken from \citet{liu14}, \citet{melendez14}, and then extended
with data from \citet{bedell14} who carefully selected lines for a high-precision abundance
determination. This data including the measured EWs are presented in the Table \ref{linelist}.

The fundamental parameters (T$_{eff}$, {log g}, [Fe/H], v$_{turb}$) of {HD 80606} and {HD 80607}
were derived by imposing excitation and ionization balance of Fe I and Fe II lines.
We used an updated version of the program FUNDPAR \citep{saffe11}, which
uses the MOOG code \citep{sneden73} together with ATLAS9 model atmospheres
\citep{kurucz93} to search the appropriate solution.
The procedure uses explicity calculated (i.e. non-interpolated) 1D LTE Kurucz's model atmospheres with
ATLAS9 and NEWODF opacities \citep{castelli-kurucz03}.

We tested the model atmospheres by using the PERL program {\it{ifconv.pl}},
which is available in the web\footnote{http://atmos.obspm.fr/index.php/documentation/7}
together with the Linux port of the Kurucz's programs.
The code checks both the convergence of the stellar flux and the flux derivative
in the ATLAS9 models, at different Rosseland optical depths.
The convergence could be a problem in the outermost layers of models calculated 
with very low T$_{eff}$ ($\sim$3500 K or less) and very low {log g}, as reported in the same page.
Under these conditions, even the LTE hypothesis probably does not hold.
However, the Kurucz's models used here are far from these values and have been tested using the
mentioned program.

The relative spectroscopic equilibrium was achieved using differential abundances
$\delta_{i}$ for each line i, defined as:

\begin{equation}
  \delta_{i} = A^{*}_{i} - A^{ref}_{i} \,,
\end{equation}

where $A^{*}_{i}$ and $A^{ref}_{i}$ are the abundances in the star of interest and in the
reference star\footnote{We use the usual abundance definition $A(X) = log(N_{X}/N_{H}) + 12$}.
The same equilibrium conditions used in \citet{saffe11} are written for the differential case as:

\begin{equation}
  s_{1} = \frac{\partial(\delta^{Fe I}_{i})} {\partial(\chi^{exc})} = 0 \,,
\end{equation}
\begin{equation}
  s_{2} = \frac{\partial(\delta^{Fe I}_{i})} {\partial(EW_{r})} = 0 \,,
\end{equation}
\begin{equation}
  D = <\delta^{Fe I}_{i}> - <\delta^{Fe II}_{i}> = 0 \,,
\end{equation}
\begin{equation}
  <\delta^{Fe I}_{i}>_{(INP)} - <\delta^{Fe I}_{i}>_{(OUT)} = 0 \,,
\end{equation}

where $\chi^{exc}$ is the excitation potential and EW$_{r}$ is the logarithm of the reduced equivalent
width. The symbol {"< >"} denote the abundance average of the different lines, while $(INP)$ and $(OUT)$
correspond to the input and output abundances in the program MOOG.
The values $s_{1}$ and $s_{2}$ are the slopes in the plots of abundance vs $\chi^{exc}$ and 
abundance vs EW$_{r}$.
In this way, equations 2 and 3 shows the independence of differential abundances with the
excitation potential and equivalent widths (by requiring null slopes $s_{1}$ and $s_{2}$),
and equation 4 is the differential equilibrium between Fe I and Fe II abundances.
Equation 5 expresses the imposed condition to the input and output abundances in the final solution.
The updated version of the program FUNDPAR searches a solution that simultaneously verifies
the conditions 2 to 5. 
The use of the 4 mentioned conditions (2 to 5) were previously tested (for the "classical"
non-differential case) using 61 main-sequence stars \citep{saffe11}, 223 giant stars \citep{jofre15}
and 9 early-type stars \citep{saffe-levato14}, obtaining very similar parameters to the literature.
Then, we applied these conditions for the differential line-by-line case, deriving for both
stars stellar parameters in agreement with the literature and with lower errors,
as we see later.

Stellar parameters of {HD 80606} and {HD 80607} were differentially determined using the Sun
as standard in a first approach, and then we recalculate the parameters of HD 80607 but using 
HD 80606 as reference. First, we determined absolute abundances for the Sun using
{5777 K} for T$_{eff}$, {4.44 dex} for {log g} and an  initial v$_{turb}$ of {1.0 km/s}. Then,
we estimated v$_{turb}$ for the Sun by the usual method of requiring zero slope in the
absolute abundances of {Fe I} lines versus EW$_{r}$ and obtained a final v$_{turb}$ of {0.91 km/s}.
We note however that the exact values are not crucial for our strictly differential study
\citep[see e.g.][]{bedell14}.

\begin{figure}
\centering
\includegraphics[width=8cm]{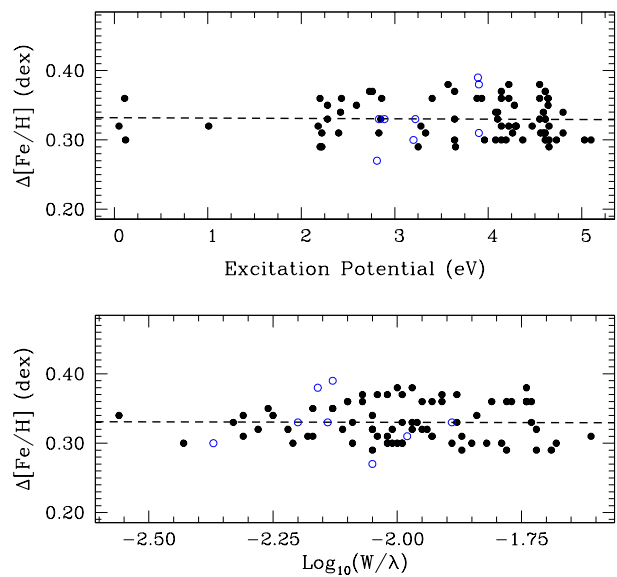}
\caption{Differential abundance vs excitation potential (upper panel) 
and differential abundance vs reduced EW (lower panel), for {HD 80606} relative to the Sun.
Filled and empty points correspond to Fe I and Fe II, respectively.
The dashed line is a linear fit to the abundance values.}
\label{equil-606-sun}%
\end{figure}

\begin{figure}
\centering
\includegraphics[width=8cm]{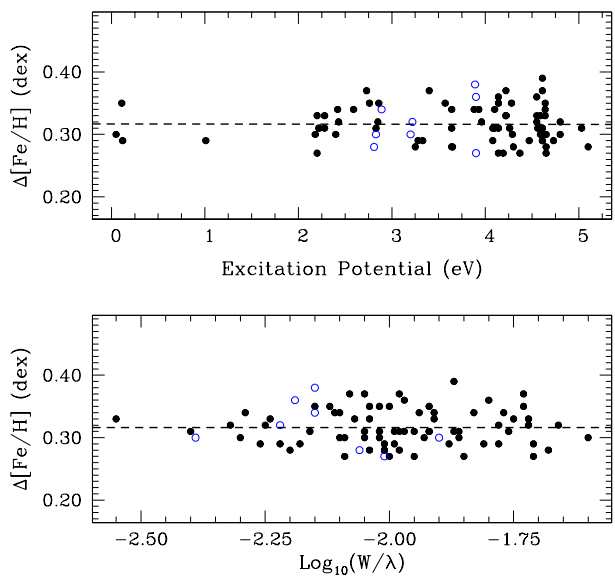}
\caption{Differential abundance vs excitation potential (upper panel) 
and differential abundance vs reduced EW (lower panel), for {HD 80607} relative to the Sun.
Filled and empty points correspond to Fe I and Fe II, respectively.
The dashed line is a linear fit to the abundance values.}
\label{equil-607-sun}%
\end{figure}

The next step was the determination of stellar parameters of {HD 80606} and {HD 80607} using the
Sun as standard. For {HD 80606} the resulting stellar parameters were {T$_{eff}$ = 5573$\pm$43 K},
{log g = 4.32$\pm$0.14 dex}, {[Fe/H] = 0.330$\pm$0.005 dex} and {v$_{turb}$ = 0.89$\pm$0.09 km/s}.
For {HD 80607} we obtained {T$_{eff}$ = 5506$\pm$21 K},
{log g = 4.31$\pm$0.11 dex}, {[Fe/H] = 0.316$\pm$0.006 dex} and {v$_{turb}$ = 0.86$\pm$0.17 km/s}.
The metallicity of the A star is slightly higher than B by 0.014 dex.
The Figures \ref{equil-606-sun} and \ref{equil-607-sun} shows the plots of abundance vs excitation
potential and abundance vs EW$_{r}$ for both stars. Filled and empty points correspond to Fe I and Fe II, 
while the dashed lines are linear fits to the differential abundance values.

The errors in the stellar parameters were derived as follows.
We estimated the change in the "observables" quantities (i.e. the slopes $s_{1}$ and $s_{2}$
and the abundance differences showed in equations 4 and 5), corresponding to individual changes
in the "measured" parameters T$_{eff}$, {log g}, [Fe/H] and v$_{turb}$ (50 K, 0.05 dex, 0.05 dex,
0.05 km/s). The mentioned changes in the "observables" are easily read in a normal execution of FUNDPAR.
A similar procedure was used previously to calculate these changes \citep[see e.g. Table 2 of][]{epstein10}.
The differences are then used to estimate the standard deviation terms which correspond
to independent parameters in the usual error propagation.
For instance, the mentioned variation of 0.05 dex in {log g} for HD 80606 produce a variation in D 
(the abundance difference between {Fe I} and {Fe II} defined in equation 4) of $\sim$0.028 dex.
Then, the individual error term in {log g} which corresponds only to the variation with D is estimated 
in a first-order approximation as $(0.05/0.028)^{2} \sigma^{2}_{D}$, where $\sigma_{D}$
is the standard deviation of the D values (estimated here using different Fe lines as
{$\sigma^{2}_{D}\simeq\sigma^{2}_{FeI} + \sigma^{2}_{FeII}$}). 
Then, we {{also}} take into account the covariance terms by using the Cauchy-Schwarz
inequality\footnote{The inequality for two variables x and y is
$\sigma^{2}_{xy} <= \sigma^{2}_{x}\sigma^{2}_{y}$ where $\sigma^{2}_{xy}$ is the mutual
covariance term and $\sigma_{x}$ , $\sigma_{y}$ are the individual dispersions.}, which
allows us to calculate the mutual covariances using the (previously calculated) individual
standard deviations. In this way, the inequality ensures that our final error adopted is
not underestimated.

\begin{figure}
\centering
\includegraphics[width=8cm]{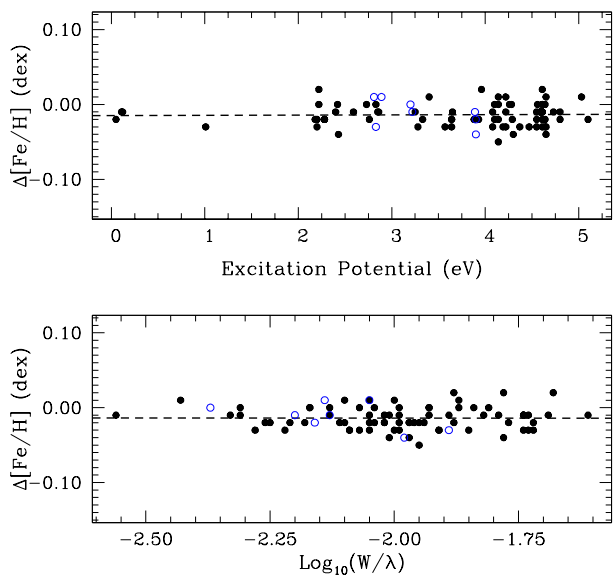}
\caption{Differential abundance vs excitation potential (upper panel) 
and differential abundance vs reduced EW (lower panel), for {HD 80607} relative to {HD 80606}.
Filled and empty points correspond to Fe I and Fe II, respectively.
The dashed line is a linear fit to the data.}
\label{equil-relat}%
\end{figure}

The process was repeated but using {HD 80606} as the reference star instead of the Sun,
fixing the parameters of the A component to perform the differential analysis.
The Figure \ref{equil-relat} shows the plots of abundance vs excitation
potential and abundance vs EW$_{r}$, using similar symbols to those used in Figures
\ref{equil-606-sun} and \ref{equil-607-sun}.
A visual inspection of the Figures \ref{equil-relat} and \ref{equil-606-sun}
shows the lower dispersion in the {HD 80607} differential abundance values using {HD 80606}
as a reference star. 
The resulting stellar parameters for {HD 80607} resulted the same as using the Sun as a
reference, but with lower dispersions: {T$_{eff}$ = 5506$\pm$14 K}, {log g = 4.31$\pm$0.08 dex},
{[Fe/H] = -0.014$\pm$0.003 dex} and {v$_{turb}$ = 0.86$\pm$0.07 km/s}. Then, the metallicity of
{HD 80607} resulted slightly lower than {HD 80606} by {0.014 dex}, equal to the
value found using the Sun as reference.

The stellar parameters derived for the A and B stars are similar to those previously
determined in the literature.
\citet{gonzalez-laws07} derived [Fe/H] = 0.349$\pm$0.073 dex for {HD 80606},
while \citet{santos04} derived (T$_{eff}$, {log g}, [Fe/H], v$_{turb}$) =
(5574$\pm$72 K, 4.46$\pm$0.20 dex, 0.32$\pm$0.09 dex, 1.14$\pm$0.09 km/s) for {HD 80606} i.e.
only {1 K} of difference compared to our result and {0.01 dex} of difference in [Fe/H].
The {log g} and v$_{turb}$ values differ by {0.14 dex} and {0.25 km/s}, respectively.
The stellar parameters derived by Santos et al. were then adopted in other works
\citep{ecuvillon06,gilli06,mortier13}.
For {HD 80607}, \citet{koleva-vazdekis12} derived {T$_{eff}$ = 5389$\pm$45 K},
{log g = 3.99$\pm$0.18 dex} and {[Fe/H] = +0.35$\pm$0.06 dex}, but adopting a fixed {v$_{turb}$ = 2.0 km/s}
for all the stars in their sample.

\begin{figure}
\centering
\includegraphics[width=8cm]{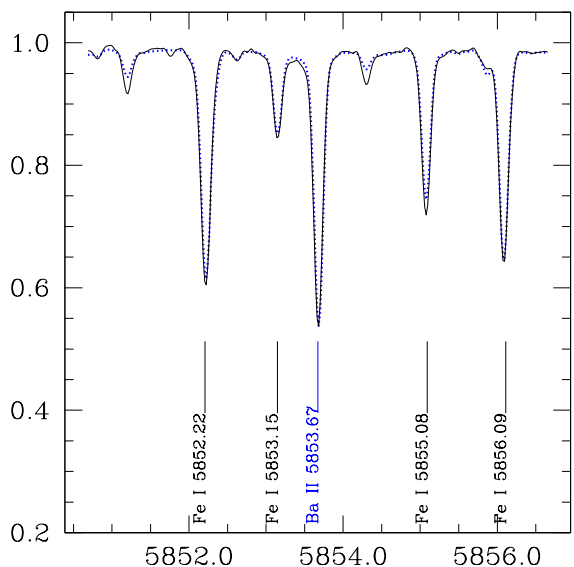}
\caption{Observed and synthethic spectra (continuous and dotted lines) near 
the line Ba II 5853.67 {\AA} for {HD 80606}. Some line identifications are showed.}
\label{606-Ba}%
\end{figure}

Once the stellar parameters of the binary components were determined
using iron lines, we computed abundances for all remaining
elements: C I, O I, Na I, Mg I, Al I, Si I, S I, Ca I, Sc I, Sc II
Ti I, Ti II, V I, Cr I, Cr II, Mn I, Fe I, Fe II, Co I, Ni I,
Cu I, Sr I, Y II and Ba II. The hyperfine structure
splitting was considered for V I, Mn I, Co I, Cu I and Ba II 
using the HFS constants of \citet{kurucz-bell95} and 
performing spectral synthesis for these species.
In the Figure \ref{606-Ba} we show an example of the observed and synthethic 
spectra in the region of the line Ba II 5853.67 {\AA} for the star {HD 80606}.
The same spectral lines were measured in both stars.
NLTE corrections were applied to the O I triplet following \citet{ramirez07}
instead of \citet{fabbian09} or \citet{takeda03}, because those works do not include corrections for {[Fe/H]$>$0}.
The NLTE abundances for O I are $\sim$0.11 dex lower than LTE values, adopting the
same correction within errors for both stars given the very similar stellar parameters.
We also applied NLTE corrections to Ba II following \citet{korotin11}, who clearly
shows that NLTE abundances are higher than LTE values for {[Fe/H]$>$0}.

In Table \ref{table.abunds} we present the final differential abundances
[X/Fe]\footnote{We used the standard notation [X/Fe] $=$ [X/H] $-$ [Fe/H]}
of {HD 80606} and {HD 80607} relative to the Sun, and the differential abundances
of {HD 80607} using {HD 80606} as the reference star.
We present both the observational errors $\sigma_{obs}$
(estimated as $\sigma/\sqrt{(n-1)}$ where $\sigma$ is the standard deviation of
the different lines) 
and systematic errors due to uncertainties in the stellar parameters
$\sigma_{par}$ (by adding quadratically the abundance variation when modifying
the stellar parameters by their uncertainties)
, as well as the total error
$\sigma_{TOT}$ obtained by adding quadratically $\sigma_{obs}$,  $\sigma_{par}$ and
the error in [Fe/H].

\begin{table*}
\centering
\caption{Differential abundances for the stars {HD 80606} and {HD 80607} relative to the Sun,
and {HD 80607} relative to {HD 80606}.
We also present the observational errors $\sigma_{obs}$, errors due to stellar
parameters $\sigma_{par}$, as well as the total error $\sigma_{TOT}$.}
\hskip -0.35in
\scriptsize
\begin{tabular}{ccccccccccccc}
\hline
\hline
  & \multicolumn{4}{c}{(HD 80606 - Sun)} &  \multicolumn{4}{c}{(HD 80607 - Sun)} & \multicolumn{4}{c}{(HD 80607 - HD 80606)}\\
Element & [X/Fe] & $\sigma_{obs}$ & $\sigma_{par}$ & $\sigma_{TOT}$ & [X/Fe]  & $\sigma_{obs}$ & $\sigma_{par}$ & $\sigma_{TOT}$ & [X/Fe]  & $\sigma_{obs}$ & $\sigma_{par}$ & $\sigma_{TOT}$  \\
\hline
{[C I/Fe]}    &    -0.040    &    0.000    &    0.057    &    0.058    &    -0.036    &    0.000    &    0.039    &    0.040    &    +0.004    &    0.000    &    0.028    &    0.028    \\
{[O I/Fe]}    &    -0.193    &    0.041    &    0.041    &    0.058    &    -0.179    &    0.057    &    0.029    &    0.064    &    +0.014    &    0.031    &    0.020    &    0.037    \\
{[Na I/Fe]}    &    -0.022    &    0.017    &    0.016    &    0.024    &    -0.048    &    0.028    &    0.011    &    0.030    &   -0.026    &    0.015    &    0.006    &    0.017    \\
{[Mg I/Fe]}    &    0.078    &    0.050    &    0.019    &    0.054    &    0.054    &    0.033    &    0.017    &    0.038    &   -0.024    &    0.021    &    0.011    &    0.024    \\
{[Al I/Fe]}    &    0.003    &    0.064    &    0.016    &    0.066    &    0.007    &    0.068    &    0.012    &    0.069    &    +0.004    &    0.007    &    0.009    &    0.012    \\
{[Si I/Fe]}    &    0.027    &    0.010    &    0.002    &    0.011    &    0.030    &    0.012    &    0.003    &    0.014    &    +0.003    &    0.004    &    0.002    &    0.005    \\
{[S I/Fe]}    &    -0.052    &    0.032    &    0.026    &    0.041    &    -0.043    &    0.050    &    0.021    &    0.055    &    +0.009    &    0.025    &    0.013    &    0.029    \\
{[Ca I/Fe]}    &    -0.048    &    0.016    &    0.015    &    0.022    &    -0.047    &    0.016    &    0.013    &    0.021    &    +0.001    &    0.003    &    0.008    &    0.009    \\
{[Sc I/Fe]}    &    0.073    &    0.035    &    0.023    &    0.043    &    0.074    &    0.041    &    0.013    &    0.043    &    +0.002    &    0.006    &    0.009    &    0.011    \\
{[Sc II/Fe]}    &    0.034    &    0.014    &    0.025    &    0.029    &    0.027    &    0.017    &    0.021    &    0.028    &   -0.007    &    0.004    &    0.015    &    0.015    \\
{[Ti I/Fe]}    &    0.033    &    0.012    &    0.009    &    0.016    &    0.042    &    0.011    &    0.009    &    0.016    &    +0.008    &    0.005    &    0.004    &    0.007    \\
{[Ti II/Fe]}    &    0.013    &    0.022    &    0.019    &    0.029    &    0.021    &    0.020    &    0.017    &    0.027    &    +0.008    &    0.014    &    0.012    &    0.019    \\
{[V I/Fe]}    &    0.085    &    0.016    &    0.013    &    0.021    &    0.091    &    0.019    &    0.013    &    0.024    &    +0.006    &    0.006    &    0.008    &    0.011    \\
{[Cr I/Fe]}    &    0.003    &    0.014    &    0.011    &    0.018    &    0.016    &    0.016    &    0.010    &    0.019    &    +0.013    &    0.005    &    0.005    &    0.008    \\
{[Cr II/Fe]}    &    0.000    &    0.054    &    0.040    &    0.067    &    0.014    &    0.070    &    0.038    &    0.080    &    +0.014    &    0.016    &    0.023    &    0.029    \\
{[Mn I/Fe]}    &    -0.023    &    0.029    &    0.023    &    0.037    &    0.014    &    0.055    &    0.027    &    0.061    &    +0.037    &    0.012    &    0.014    &    0.019    \\
{[Co I/Fe]}    &    0.191    &    0.020    &    0.016    &    0.027    &    0.231    &    0.024    &    0.016    &    0.030    &    +0.040    &    0.008    &    0.010    &    0.013    \\
{[Ni I/Fe]}    &    0.078    &    0.007    &    0.004    &    0.009    &    0.078    &    0.007    &    0.005    &    0.010    &   -0.001    &    0.004    &    0.003    &    0.005    \\
{[Cu I/Fe]}    &    -0.070    &    0.050    &    0.040    &    0.064    &    -0.086    &    0.060    &    0.040    &    0.072    &   -0.016    &    0.020    &    0.025    &    0.032    \\
{[Sr I/Fe]}    &    0.120    &    0.050    &    0.086    &    0.100    &    0.094    &    0.060    &    0.106    &    0.120    &   -0.026    &    0.020    &    0.055    &    0.058    \\
{[Y II/Fe]}    &    -0.002    &    0.028    &    0.034    &    0.044    &    0.030    &    0.027    &    0.042    &    0.050    &    +0.032    &    0.009    &    0.025    &    0.026    \\
{[Ba II/Fe]}    &    0.190    &    0.050    &    0.040    &    0.064    &    0.177    &    0.060    &    0.040    &    0.072    &   -0.013    &    0.020    &    0.025    &    0.032    \\
\hline
\end{tabular}
\normalsize
%\tablebib{R1 \citep{abt79}, R2 \citep{mermilliod83}, R3 \citep{shore87}}
\label{table.abunds}
\end{table*}

\section{Results and discussion}

\begin{figure}
\centering
\includegraphics[width=8cm]{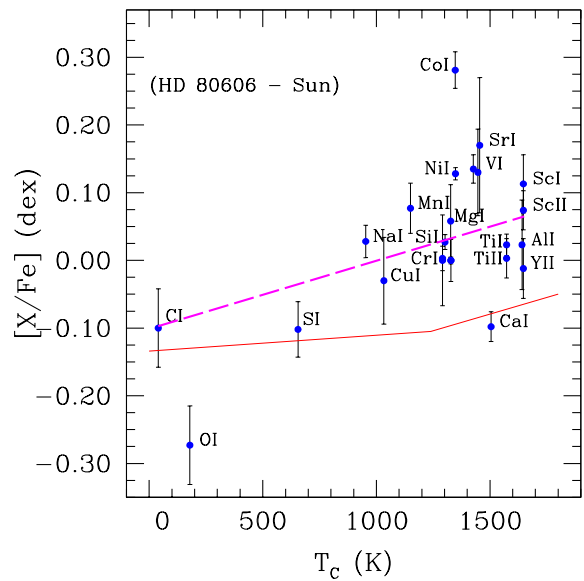}
\caption{Differential abundances {(HD 80606 - Sun)} vs condensation temperature T$_{c}$.
The dashed line is a weighted linear fit to the differential abundance values,
while the continuous line shows the solar-twins trend of \citet{melendez09}.}
\label{abund-606-sun}%
\end{figure}

\begin{figure}
\centering
\includegraphics[width=8cm]{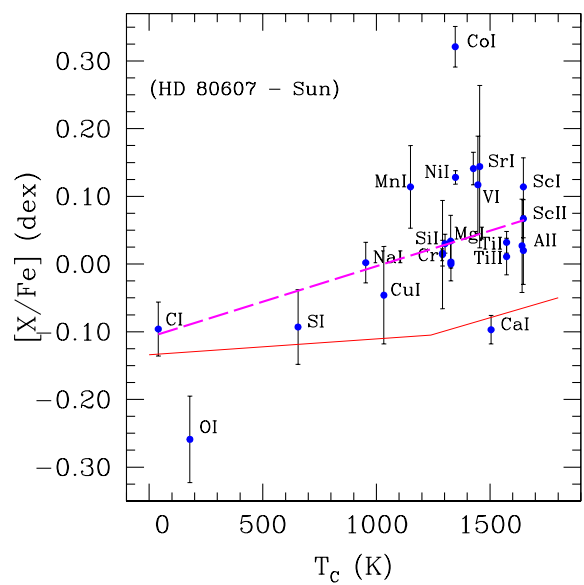}
\caption{Differential abundances {(HD 80607 - Sun)} vs condensation temperature T$_{c}$.
The dashed line is a weighted linear fit to the differential abundance values,
while the continuous line shows the solar-twins trend of \citet{melendez09}.}
\label{abund-607-sun}%
\end{figure}

We present in the Figures \ref{abund-606-sun} and \ref{abund-607-sun} the differential abundances
of {HD 80606} and {HD 80607} relative to the Sun. The condensation temperatures were taken from the
50\% T$_{c}$ values derived by \citet{lodders03}. 
The individual comparison between one component (e.g. HD 80606) and the Sun, is possibly
affected by Galactic Chemical Evolution (GCE) effects, due to their different chemical natal
environments \citep[see e.g.][and references therein]{tayo14,molla15}.
On the other hand, supossing that the stars of the binary system born at the same place/time,
we discard GCE effects when comparing differentially the components between them,
which is an important advantage of this method.
Then, we corrected by GCE effects (only when comparing star-Sun) by adopting the fitting trends
of \citet{gonzhern13} (see their Figure 2, the plots of [X/Fe] vs [Fe/H]) to derive the values of
[X/Fe] at [Fe/H]$\sim$0.32 dex. A similar procedure was previously used by \citet{liu14}
to correct by GCE the abundances in the binary system {HAT-P-1}.
Filled points in the Figures \ref{abund-606-sun} and \ref{abund-607-sun} correspond to the
differential abundances for the stars {HD 80606} and {HD 80607}, respectively.
For reference, we also included in these Figures the solar-twins trend of M09 using a continuous line,
vertically shifted to compare the slopes.
We included a weighted linear fit\footnote{We used as weight the inverse of the total abundance
error $\sigma_{TOT}$.} to all abundance values, showed with dashed lines in the Figures
\ref{abund-606-sun} and \ref{abund-607-sun}. It is interesting to note that the slopes of the
linear fits are similar to the trend of the solar-twins of M09 for the refractory elements.

In the Figures \ref{abund-606-sun} and \ref{abund-607-sun}, 
the abundance of {O I} presents a low value compared to other volatile elements,
while the abundances of {Co I} and {Ca I} seem to deviate
from the general trend of the refractory elements
(see also the next Figures \ref{refr-606-sun} and \ref{refr-607-sun}).
For both stars, we derived the {O I} abundance by measuring
EWs of the {O I} triplet at 7771 \AA\ and applied NLTE corrections following \citet{ramirez07}.
As we noted previously, the NLTE corrections decrease the abundance in $\sim$0.11 dex,
However, even the LTE values seem to be relatively low; we do not find a clear reason for this.
The forbidden [O I] lines at 6300.31 \AA\ and 6363.77 \AA\ are weak and slightly asymetric
in our stars. Both [O I] lines are blended in the solar spectra: with two {N I} lines in the red
wing of [O I] 6300.31 \AA\ and with CN near [O I] 6363.77 \AA\ \citep{lambert78,johansson03,bensby04}.
Then, we prefer to avoid these weak [O I] lines in our calculation and use only the O I triplet.
For the case of {Co I}, we take into account the HFS in the abundance calculation,
however NLTE effects could also play a role in the {Co I} lines of solar-type stars
\citep[see e.g.][]{bergemman08,bergemman10}. \citet{mashonkina07} studied NLTE effects
in the {Ca I} lines of late-type stars, and derived higher NLTE abundances than in LTE
for most {Ca I} lines, using a model with {T$_{eff}$ = 5500 K} and {[Fe/H] = 0}.
For these stellar parameters the corrections amount up to 0.08 dex, with an average of
$\sim$0.05 dex. However, we caution that these studies for {Co I} and {Ca I} do not
include corrections for stars with {[Fe/H]$>$0}. Therefore, we excluded these species
(O I, Co I and Ca I) from the calculation of the linear fits.

\citet[][ hereafter R10]{ramirez10} studied the abundance results from six different abundance
surveys and verified the findings of M09 about the T$_{c}$ trends in the Sun and the
terrestrial planet formation signature.
They studied the possible dependence of the T$_{c}$ trends with [Fe/H] using in particular
the sample of \citet[][ hereafter N09]{neves09}. The authors showed that the "solar anomaly" 
(i.e. the T$_{c}$ trend for the refractory elements in the Sun)
is also observed comparing the Sun with solar-analogs at both
{[Fe/H]$\sim$-0.2 dex} and {[Fe/H]$\sim$0.0 dex}.
However, for an average metallicity of {[Fe/H]$\sim$+0.2 dex}, the solar analogs from
N09 shows a T$_{c}$ trend for refractories similar to the Sun (see e.g. their Figure 7).
R10 interpret this result suggesting that, at high metallicity values,
the probability of stars with and without T$_{c}$ trends should be similar, and then,
in average, no general trend with T$_{c}$ result for the refractory elements.
The authors also propose that it may be possible to distinguish metal-rich stars
that show and do not show the planet formation signature from the T$_{c}$ slopes
of the refractory elements. Then, given that {HD 80606} and {HD 80607} present high
metallicity values, it seems reasonable also a comparison of the refractories with the
solar-analog stars with {[Fe/H]$\sim$+0.2 dex} from N09.

The differential abundances of the refractory species are showed in the Figures
\ref{refr-606-sun} and \ref{refr-607-sun}.
We include in these Figures the trend of the solar-analog stars with {[Fe/H]$\sim$+0.2 dex}
from N09 using a short-dased line, which shows almost an horizontal tendence.
The solar-twins T$_{c}$ trend of M09 is also showed with a continuous line.
The tendences of N09 and M09 are vertically shifted for comparison.
A weighted linear fit to the refractory species of {HD 80606} and {HD 80607}
is presented with a long-dashed line.
The refractory elements does not seem to follow an horizontal trend such as the sample of N09.
The general trend of refractory species for both {HD 80606} and {HD 80607},
are more similar to the solar-twins of M09 than to the solar-analogs stars
with {[Fe/H]$\sim$+0.2 dex} from N09.
The Sun is depleted in refractory elements compared to the solar-twins of M09,
however the solar-analogs with {[Fe/H]$\sim$+0.2 dex} from N09 present a similar T$_{c}$ trend
compared to the Sun, as showed by R10.
Then, following a reasoning similar to M09 and R10,
the stars {HD 80606} and {HD 80607} do not seem to be depleted in refractory elements
with respect to solar twins, which is different for the case of the Sun.
In other words, the terrestrial planet formation would have been
less efficient in the stars of this binary system than in the Sun.

\begin{figure}
\centering
\includegraphics[width=8cm]{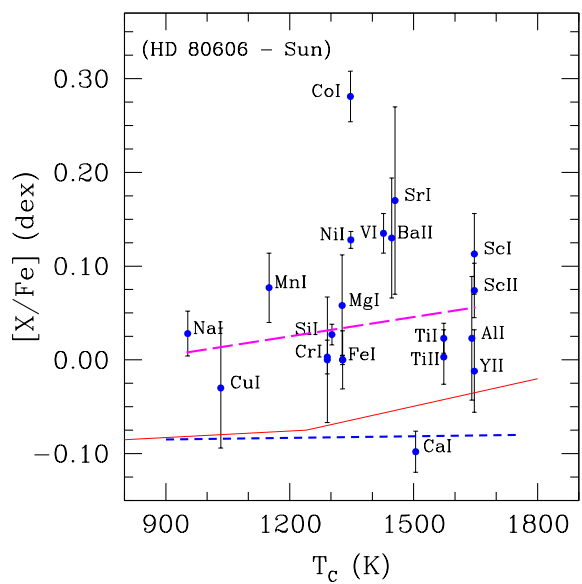}
\caption{Differential abundances {(HD 80606 - Sun)} vs condensation temperature T$_{c}$
for the refractory elements.
The long-dashed line shows a weighted linear fit to the abundance values.
The continuous and short-dashed lines correspond to the solar-twins trend of M09,
and the solar-analogs with {[Fe/H]$\sim$+0.2 dex} from N09.}
\label{refr-606-sun}%
\end{figure}

\begin{figure}
\centering
\includegraphics[width=8cm]{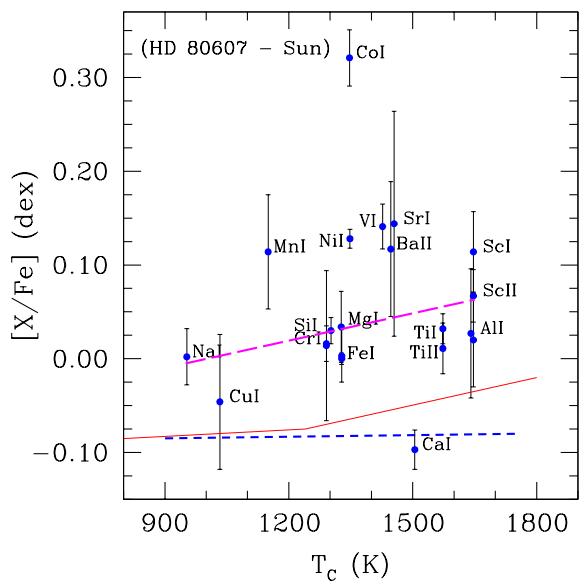}
\caption{Differential abundances {(HD 80607 - Sun)} vs condensation temperature T$_{c}$
for the refractory elements.
The symbols are the same of Figure \ref{refr-606-sun}.}
\label{refr-607-sun}%
\end{figure}

\begin{figure}
\centering
\includegraphics[width=8cm]{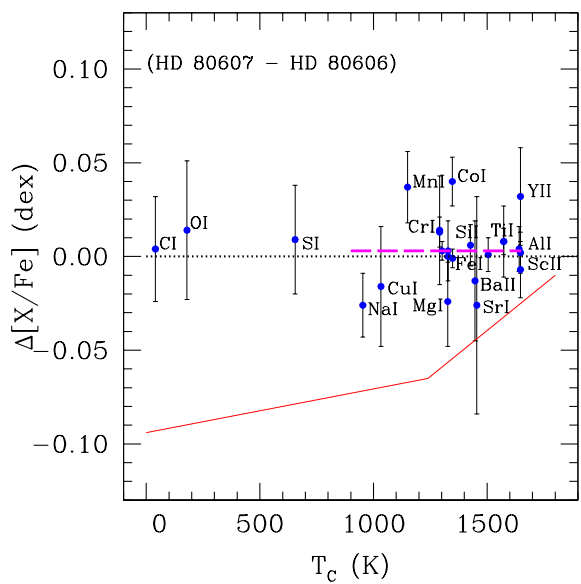}
\caption{Differential abundances (HD 80607 - HD 80606) vs condensation temperature T$_{c}$.
The long-dashed line is a weighted linear fit to the refractory species.
The solar-twins trend of \citet{melendez09} is showed with a continuous line.
The horizontal line at 0.0 dex is included for reference.}
\label{relat.tc}%
\end{figure}

The line-by-line differential abundances between {HD 80606} and {HD 80607}
greatly diminishes the errors in the calculation and GCE effects in the results,
due to their remarkably similar stellar parameters and due to the same (initial)
chemical composition.
In the Figure \ref{relat.tc} we show the differential abundances of {HD 80607} vs T$_{c}$
but using {HD 80606} as the reference star. %In the Figure \ref{relat.z} we show differential abundances (in this case as [X/H]) vs atomic number z, including a dotted line at 0.0 dex which corresponds to the reference star {HD 80607}. In the Figure \ref{relat.tc} we present abundances vs T$_{c}$.
The continuous line in this Figure presents the solar-twins trend of M09 (vertically shifted),
while the long-dashed line is a weighted linear fit to the refractory elements.
We included an horizontal line at 0.0 dex for reference. 

%\begin{figure}
%\centering
%\includegraphics[width=8cm]{sm4.eps}
%\caption{Differential abundances {(HD 80606 - HD 80607)} vs atomic number z.
%The horizontal line corresponds to the reference star {HD 80607}.}
%\label{relat.z}%
%\end{figure}

Most elements present slightly higher
abundance values in {HD 80606} compared to {HD 80607}, with an average difference of
{+0.010$\pm$0.019 dex}. In particular, the difference for the Fe I abundances is
{+0.014$\pm$0.003 dex} i.e. {HD 80606} slightly more metal-rich than {HD 80607}.
From the Figure \ref{relat.tc}, the abundances of the volatile does not
seem to be different from the refractory elements.
Their average abundances are -0.005$\pm$0.005 dex and -0.011$\pm$0.005 dex
i.e. almost the same within the errors.
In the Figure \ref{relat.tc}, the slope of the differential abundances 
is {-1.20$\pm$16.5 10$^{-6}$ dex/K} for the refractory elements.
For comparison, the slope of refractories between the components of the binary system
{16 Cyg} resulted {1.88$\pm$0.79 10$^{-5}$ dex/K} and showing clearly a higher abundance in
refractory than volatile elements \citep{tuccimaia14}.
Then, although {HD 80606} seems to present a slightly higher Fe I abundance
than {HD 80607}, there is no clear difference between refractory and volatile
elements nor a significative trend with T$_{c}$.
This would imply that there is no clear evidence of terrestrial planet formation
in this binary system.
Similarly, \citet{liu14} did not find a trend with T$_{c}$ in the binary system
{HAT-P-1} and concluded that the presence of a giant planet does not neccesarily
introduce a chemical signature in their host stars.
This is in line with some previous literature works, who propose that the
presence of close-in giant planets might prevent the formation of terrestrial planets
\citep{melendez09,steffen12}.
For the case of eccentric giant planets, numerical simulations also found that the
early dynamical evolution of giant planets clear out most of the terrestrial planets
in the inner zone \citep{veras-armitage05,veras-armitage06,raymond11}.

\subsection{A planet around HD 80607?}

Up to now, there is no planet detected around HD 80607.
The photometry of HD 80607 is relatively flat i.e. a transit-like event
is not observed \citep{fossey09,pont10}.
To our knowledge, this object is not included in the current radial velocity
surveys.

However, given the abundance results of this study and the confirmed presence
of a giant planet (with very high eccentricity) only around {HD 80606}, we can
speculate about a possible planet formation scenario in this binary system.
The occurrence of planets was fit by \citet{fischer-valenti05} using a power law
as a function of the metallicity:
{P $=$ 0.03 (N$_{Fe}$/N$_{H}$)$^{2}$/(N$_{Fe}$/N$_{H}$)$^{2}_{\sun}$}.
Then, the probability increases by a factor of 5 when the Fe abundance increase from
{[Fe/H] = 0.0 dex} to {[Fe/H] = 0.3 dex}. This high probability together with the fact
that {HD 80606} already host a giant planet, and given the very similar stellar parameters
with {HD 80607}, suggest that the giant planet formation process in {HD 80607}
could be also a very plausible hypothesis.
Possibly, the metals missing in {HD 80607} compared to {HD 80606} have been used
to form this (hypothetic) giant planet. \citet{tuccimaia14} make a similar
suggestion to explain the slightly different metallicities between the
components of the binary system {16 Cyg}. Moreover, there are binary systems where
each component hosts a planet and the metallicity resulted slightly different
between their stars, such as in the system {XO-2} \citep{damasso15}.
Then, probably due to the mutual interactions in this binary system, 
{HD 80606} resulted with one of the most eccentric planets to date
\citep[see e.g.][]{wu-murray03}, while the {HD 80607} system may have had 
its giant planet ejected.
In fact, the possible companion around {HD 80607} could be an ejected or maybe
an undetected (such as a long period) planet.
We stress, however, that this is only a speculative comment and should be taken
with caution.

Previous works showed that the global frequency of planets in wide binaries is not
statistically different from that of planets in single stars, with no significant
dependence of the binary separation \citep{bonavita-desidera07}. 
Also, the properties of planets in wide binaries are compatible with those of planets
orbiting single stars, except for a possible increase of high-excentricity planets
\citep{desidera-barbieri07}.
However, the presence of closer stellar companions with separation 100-300 AU could
modify the evolution of giant planets around binary components \citep{desidera-barbieri07}.

More recently, \citet{wang15} studied 84 KOIs (Kepler Object of Interest)
with al least one gas giant planet detected within 1 AU and a control sample of
field stars in the solar neighborhood. The authors found a dependence
of the stellar multiplicity rate (MR) as a function of the stellar separation a.
They derived MRs of $\sim$0\%, $\sim$34\% and $>$34\% for binary separations of
a $<$ 20 AU, {{20 AU} $<$ a $<$ {200 AU}}, and a $>$ 200 AU, respectively.
In other words, no stellar companion has been found within 20 AU for Kepler stars with
gas giant planets, while gas giant planet formation is not significantly affected
by stellar companions beyond 200 AU.
Then, this work shows that the binary separation plays a role in close binaries rather
than in wide binaries, such as HD 80606 (a $\sim$1000 AU).
This is in agreement with \citet{zuckerman14}, who found that the the presence of a wide
stellar companion (a $\geq$ 1000 AU) does not diminish the likelihood of a wide-orbit
planetary system.

\citet{wang15} also studied the possible physical differences between the components
of binaries hosting planets. They suggest that the stellar companions of host stars with
a planet period P$>$70 d tend to be fainter than the shorter-period counterparts. However,
they caution that this apparent effect may be due to a lack of sensivity for fainter
stellar companions and suggest more follow-up observations to support or disprove it.

Using numerical simulations, \citet{wu-murray03} suggest that the high exccentricity
of the planet {HD 80606 b} is probably due to the influence of the companion HD 80607
through a Kozai mechanism\footnote{The Kozai mechanism are oscilations in the eccentricity
and inclination  of a planet due to the presence of a remote stellar companion, see e.g.
\citet{kozai62}.} combined with a tidal dissipation.
On the other hand, \citet{kaib13} showed a possible variable nature of wide binaries
due to the Milky Way tidal field, including a reshape of their planetary systems.
In this scenario, they obtained an instability fraction (i.e. number of planetary
ejections within 10 Gyr of evolution) depending on the binary's mass and separation.
Using the binary parameters of HD 80606, they obtained a fraction $\sim$ 50\%
(see their Fig. 2). Although these simulations do not include the possibility of
a planet around HD 80607, they showed that the planetary configuration in this
binary system could be strongly affected, and the possible ejection of a planet 
could not be totally ruled out.

\section{Conclusions}

Following the aims of this study, we performed a high-precision differential abundance
determination in both components of the remarkable binary system {HD 80606 - HD 80607},
in order to possibly detect a signature of terrestrial planet formation.
Both stars present very similar stellar parameters, which greatly diminishes
the errors in the abundance determination and GCE effects.
The star {HD 80606} hosts a giant (high-eccentricity) planet while there is no planet
detected around {HD 80607}.
First, we derived stellar parameters and differential abundances of both stars using the
Sun as the reference star.
We compared the possible temperature condensation T$_{c}$ trends of the stars
with the solar-twins trend of \citet{melendez09} and then with a sample
of solar-analog stars with {[Fe/H]$\sim$+0.2 dex} from \citet{neves09}.
Our calculation included NLTE corrections for O I and Ba II as well as GCE
corrections for all chemical species.
From these comparisons, we concluded that the stars {HD 80606} and {HD 80607} do not
seem to be depleted in refractory elements, different to the case of the Sun
\citep{melendez09}. In other words, the terrestrial planet formation would
have been less efficient in the stars of this binary system than in the Sun.

Then, we also compared differentially {HD 80607} but using {HD 80606} as the reference star.
{HD 80606} resulted slightly more metal-rich than {HD 80607} by {+0.014$\pm$0.003 dex}.
However, we do not find a clear difference between refractory and volatile
elements nor a significative trend with T$_{c}$ between both stars.
{{In comparing the stars to each other, the lack of a trend for refractory elements with
T$_{c}$ implies that the presence of a giant planet does not necessarily imprint a chemical
signature on its host star,}} similar to the result of \citet{liu14} for the
binary system {HAT-P-1}. This is in agreement with \citet{melendez09},
who suggest that the presence of close-in giant planets might prevent the
formation of terrestrial planets.
Finally, we speculate about a possible (ejected or non-detected) planet
around {HD 80607}.
We strongly encourage high-precision abundance studies in binary systems
with similar components, which is a crucial tool for helping to detect the
possible chemical pattern of the planet formation process.

\begin{acknowledgements}
We thank the anonymous referee for their constructive comments
that greatly improved the paper.
The authors wish to recognize and acknowledge the very significant cultural role
and reverence that the summit of Mauna Kea has always had within the indigenous
Hawaiian community.  We are most fortunate to have the opportunity to conduct
observations from this mountain. 
The authors also thank Drs. R. Kurucz and C. Sneden for making their codes available to them.
\end{acknowledgements}

\onltab{1}{
\onecolumn
\begin{longtable}{ccccccc}
\caption{Line list used in this work. The columns present the element, 
wavelength $\lambda$, Excitation Potential EP, log gf, Equivalent Widths
of HD 80606, HD 80607 and Sun ($EW_{1}$, $EW_{2}$ and $EW_{Sun}$).
The abundances of lines without EWs are measured using synthetic spectra.}
\\
\hline \hline
Element  & $\lambda$  & EP   & log gf & $EW_{1}$ & $EW_{2}$ & $EW_{Sun}$ \\
         & [\AA]      & [eV] & [dex]  & [m\AA]   & [m\AA]   & [m\AA]     \\
\hline
\endfirsthead
\caption{Continued.}\\
\hline \hline
Element  & $\lambda$  & EP   & log gf & $EW_{1}$ & $EW_{2}$ & $EW_{Sun}$ \\
         & [\AA]      & [eV] & [dex]  & [m\AA]   & [m\AA]   & [m\AA]     \\
\hline
\endhead
\hline
\endfoot
    6.00 &   5052.167 &    7.680 &   -1.240 &    42.0 &    39.4 &    33.6 \\
    6.00 &   6587.610 &    8.540 &   -1.050 &    16.6 &    14.8 &    12.1 \\
    8.00 &   7771.944 &    9.150 &    0.370 &    65.7 &    58.4 &    66.9 \\
    8.00 &   7774.166 &    9.150 &    0.220 &    61.5 &    59.3 &    62.1 \\
    8.00 &   7775.388 &    9.150 &    0.000 &    49.8 &    47.2 &    45.0 \\
   11.00 &   4751.822 &    2.100 &   -2.080 &    36.7 &    38.5 &    15.7 \\
   11.00 &   5148.838 &    2.100 &   -2.040 &    31.0 &    31.7 &    13.8 \\
   11.00 &   6154.225 &    2.100 &   -1.550 &    75.3 &    80.6 &    39.2 \\
   11.00 &   6160.747 &    2.100 &   -1.250 &    91.8 &    94.1 &    56.9 \\
   12.00 &   4730.040 &    4.340 &   -2.390 &   112.1 &   108.7 &    68.6 \\
   12.00 &   5711.088 &    4.340 &   -1.730 &   147.0 &   144.6 &   106.6 \\
   12.00 &   6318.717 &    5.110 &   -1.950 &    62.1 &    63.8 &    37.3 \\
   12.00 &   6319.236 &    5.110 &   -2.160 &    52.1 &    51.5 &    24.2 \\
   13.00 &   5557.070 &    3.140 &   -2.210 &    25.5 &    25.6 &    13.4 \\
   13.00 &   6696.018 &    3.140 &   -1.480 &    62.3 &    64.7 &    36.0 \\
   13.00 &   6698.667 &    3.140 &   -1.780 &    47.0 &    48.6 &    20.8 \\
   14.00 &   5488.983 &    5.610 &   -1.690 &    38.1 &    36.8 &    18.5 \\
   14.00 &   5517.540 &    5.080 &   -2.500 &    24.6 &    23.6 &    12.2 \\
   14.00 &   5645.611 &    4.930 &   -2.040 &    56.7 &    56.7 &    35.8 \\
   14.00 &   5665.554 &    4.920 &   -1.940 &    65.0 &    65.5 &    39.3 \\
   14.00 &   5684.484 &    4.950 &   -1.550 &    81.2 &    80.6 &    61.0 \\
   14.00 &   5690.425 &    4.930 &   -1.770 &    67.7 &    67.6 &    48.5 \\
   14.00 &   5701.104 &    4.930 &   -1.950 &    58.6 &    56.1 &    40.3 \\
   14.00 &   5753.640 &    5.620 &   -1.330 &    71.6 &    72.7 &    43.5 \\
   14.00 &   5772.145 &    5.082 &   -1.653 &    74.1 &    74.7 &    51.8 \\
   14.00 &   5793.073 &    4.930 &   -1.960 &    64.7 &    62.2 &    42.9 \\
   14.00 &   5948.540 &    5.080 &   -1.208 &   108.8 &   108.3 &    84.4 \\
   14.00 &   6125.021 &    5.610 &   -1.500 &    51.2 &    49.6 &    31.7 \\
   14.00 &   6145.015 &    5.620 &   -1.410 &    59.4 &    58.8 &    38.7 \\
   14.00 &   6195.460 &    5.870 &   -1.666 &    33.8 &    34.2 &    15.2 \\
   14.00 &   6243.823 &    5.620 &   -1.270 &    61.8 &    59.2 &    43.9 \\
   14.00 &   6244.476 &    5.620 &   -1.320 &    71.4 &    70.4 &    45.4 \\
   14.00 &   6741.630 &    5.980 &   -1.650 &    28.6 &    27.8 &    15.2 \\
   14.00 &   7034.903 &    5.870 &   -0.780 &    81.6 &    82.4 &    62.8 \\
   14.00 &   7405.770 &    5.614 &   -0.720 &   112.3 &   111.8 &    88.7 \\
   16.00 &   4695.443 &    6.530 &   -1.830 &    12.0 &    12.6 &     8.2 \\
   16.00 &   6046.000 &    7.870 &   -0.150 &    28.4 &    24.8 &    20.3 \\
   16.00 &   6052.656 &    7.870 &   -0.400 &    17.8 &    16.9 &    13.2 \\
   16.00 &   6743.540 &    7.870 &   -0.600 &    12.6 &    10.8 &     9.7 \\
   20.00 &   5260.387 &    2.520 &   -1.720 &    52.2 &    54.1 &    32.5 \\
   20.00 &   5261.710 &    2.520 &   -0.680 &   127.8 &   131.5 &   100.6 \\
   20.00 &   5512.980 &    2.930 &   -0.460 &   114.2 &   116.0 &    83.8 \\
   20.00 &   5590.114 &    2.520 &   -0.570 &   110.9 &   113.6 &    92.8 \\
   20.00 &   5867.562 &    2.930 &   -1.570 &    42.5 &    43.5 &    23.5 \\
   20.00 &   6156.020 &    2.520 &   -2.497 &    19.8 &    19.8 &     8.7 \\
   20.00 &   6161.297 &    2.520 &   -1.270 &    82.9 &    84.9 &    59.5 \\
   20.00 &   6166.439 &    2.520 &   -1.140 &    94.1 &    95.8 &    69.6 \\
   20.00 &   6169.550 &    2.520 &   -0.580 &   139.9 &   141.8 &   108.7 \\
   20.00 &   6455.598 &    2.520 &   -1.340 &    80.3 &    82.6 &    55.2 \\
   20.00 &   6471.662 &    2.530 &   -0.690 &   112.6 &   115.6 &    91.0 \\
   20.00 &   6499.650 &    2.520 &   -0.820 &   102.9 &   105.2 &    85.5 \\
   21.00 &   4743.821 &    1.450 &    0.350 &    24.9 &    26.8 &     9.2 \\
   21.00 &   5081.570 &    1.450 &    0.300 &    25.0 &    27.3 &     7.4 \\
   21.00 &   5520.497 &    1.860 &    0.550 &    16.9 &    17.9 &     6.1 \\
   21.00 &   5671.821 &    1.450 &    0.550 &    40.2 &    42.6 &    14.7 \\
   21.10 &   5657.870 &    1.510 &   -0.300 &    77.6 &    75.6 &    65.7 \\
   21.10 &   5669.055 &    1.500 &   -1.200 &    48.6 &    47.4 &    36.4 \\
   21.10 &   5684.190 &    1.510 &   -0.950 &    50.9 &    49.9 &    37.7 \\
   21.10 &   6245.630 &    1.510 &   -1.030 &    49.2 &    48.6 &    35.2 \\
   21.10 &   6279.760 &    1.500 &   -1.200 &    42.8 &    40.9 &    30.1 \\
   21.10 &   6320.843 &    1.500 &   -1.850 &    14.3 &    13.7 &     7.6 \\
   21.10 &   6604.578 &    1.360 &   -1.150 &    52.5 &    52.0 &    35.5 \\
   22.00 &   4617.280 &    1.750 &    0.450 &    85.6 &    88.1 &    64.1 \\
   22.00 &   4645.190 &    1.730 &   -0.670 &    40.8 &    44.8 &    21.7 \\
   22.00 &   4656.470 &    0.000 &   -1.310 &    91.4 &    95.6 &    68.4 \\
   22.00 &   4758.120 &    2.250 &    0.430 &    61.4 &    63.2 &    43.0 \\
   22.00 &   4759.272 &    2.260 &    0.510 &    66.3 &    67.6 &    47.0 \\
   22.00 &   4778.258 &    2.240 &   -0.220 &    35.3 &    35.3 &    15.4 \\
   22.00 &   4820.410 &    1.500 &   -0.440 &    66.7 &    69.2 &    43.0 \\
   22.00 &   4999.500 &    0.830 &    0.270 &   135.8 &   138.4 &   104.5 \\
   22.00 &   5022.871 &    0.830 &   -0.430 &    93.7 &    96.4 &    70.9 \\
   22.00 &   5024.850 &    0.820 &   -0.560 &    98.5 &   100.7 &    70.0 \\
   22.00 &   5039.960 &    0.020 &   -1.200 &   100.7 &   101.0 &    76.2 \\
   22.00 &   5071.490 &    1.460 &   -0.800 &    57.0 &    57.2 &    27.7 \\
   22.00 &   5147.479 &    0.000 &   -2.010 &    59.9 &    64.1 &    34.1 \\
   22.00 &   5219.700 &    0.020 &   -2.240 &    59.7 &    64.5 &    29.1 \\
   22.00 &   5471.200 &    1.440 &   -1.400 &    21.9 &    25.2 &     7.9 \\
   22.00 &   5490.150 &    1.460 &   -0.930 &    42.7 &    46.7 &    21.0 \\
   22.00 &   5689.459 &    2.300 &   -0.360 &    29.0 &    31.1 &    11.5 \\
   22.00 &   5739.464 &    2.250 &   -0.600 &    20.2 &    22.1 &     6.3 \\
   22.00 &   5766.330 &    3.290 &    0.326 &    22.3 &    23.1 &     9.0 \\
   22.00 &   5866.452 &    1.070 &   -0.840 &    76.7 &    79.6 &    47.6 \\
   22.00 &   6064.630 &    1.050 &   -1.959 &    25.6 &    27.1 &     7.8 \\
   22.00 &   6091.174 &    2.270 &   -0.420 &    35.5 &    37.9 &    14.7 \\
   22.00 &   6126.217 &    1.070 &   -1.420 &    46.2 &    49.5 &    22.4 \\
   22.00 &   6258.104 &    1.440 &   -0.350 &    79.2 &    82.3 &    50.4 \\
   22.00 &   6303.753 &    1.443 &   -1.509 &    24.5 &    25.7 &     8.0 \\
   22.00 &   6312.234 &    1.460 &   -1.496 &    20.5 &    23.6 &     6.8 \\
   22.00 &   6599.104 &    0.900 &   -2.029 &    27.2 &    29.6 &     8.8 \\
   22.00 &   6743.130 &    0.899 &   -1.630 &    43.2 &    47.1 &    17.8 \\
   22.00 &   7949.150 &    1.500 &   -1.456 &    29.4 &    32.2 &     8.2 \\
   22.10 &   4636.330 &    1.160 &   -3.150 &    30.5 &    27.3 &    17.6 \\
   22.10 &   4779.985 &    2.048 &   -1.260 &    72.2 &    76.3 &    65.2 \\
   22.10 &   4798.532 &    1.080 &   -2.670 &    53.2 &    52.4 &    42.6 \\
   22.10 &   4865.611 &    1.120 &   -2.810 &    55.9 &    54.2 &    40.7 \\
   22.10 &   4911.193 &    3.120 &   -0.540 &    64.1 &    64.1 &    53.3 \\
   22.10 &   5005.160 &    1.570 &   -2.720 &    31.0 &    31.5 &    19.6 \\
   22.10 &   5418.767 &    1.580 &   -2.110 &    60.5 &    59.1 &    49.4 \\
   23.00 &   4875.442 &    0.040 &   -3.375 \\
   23.00 &   4875.454 &    0.040 &   -2.260 \\
   23.00 &   4875.461 &    0.040 &   -2.964 \\
   23.00 &   4875.468 &    0.040 &   -1.420 \\
   23.00 &   4875.471 &    0.040 &   -2.064 \\
   23.00 &   4875.477 &    0.040 &   -2.742 \\
   23.00 &   4875.483 &    0.040 &   -1.561 \\
   23.00 &   4875.485 &    0.040 &   -2.010 \\
   23.00 &   4875.491 &    0.040 &   -2.617 \\
   23.00 &   4875.495 &    0.040 &   -1.725 \\
   23.00 &   4875.497 &    0.040 &   -2.032 \\
   23.00 &   4875.502 &    0.040 &   -2.566 \\
   23.00 &   4875.505 &    0.040 &   -1.923 \\
   23.00 &   4875.506 &    0.040 &   -2.123 \\
   23.00 &   4875.509 &    0.040 &   -2.596 \\
   23.00 &   4875.511 &    0.040 &   -2.178 \\
   23.00 &   4875.511 &    0.040 &   -2.311 \\
   23.00 &   4875.515 &    0.040 &   -2.566 \\
   23.00 &   5703.555 &    1.050 &   -0.777 \\
   23.00 &   5703.569 &    1.050 &   -0.993 \\
   23.00 &   5703.569 &    1.050 &   -1.403 \\
   23.00 &   5703.580 &    1.050 &   -1.242 \\
   23.00 &   5703.580 &    1.050 &   -1.276 \\
   23.00 &   5703.581 &    1.050 &   -2.268 \\
   23.00 &   5703.589 &    1.050 &   -1.250 \\
   23.00 &   5703.589 &    1.050 &   -1.715 \\
   23.00 &   5703.590 &    1.050 &   -1.840 \\
   23.00 &   5703.596 &    1.050 &   -1.414 \\
   23.00 &   5703.596 &    1.050 &   -1.590 \\
   23.00 &   5703.601 &    1.050 &   -1.414 \\
   23.00 &   5727.008 &    1.080 &   -0.693 \\
   23.00 &   5727.016 &    1.080 &   -1.701 \\
   23.00 &   5727.022 &    1.080 &   -3.003 \\
   23.00 &   5727.028 &    1.080 &   -0.798 \\
   23.00 &   5727.035 &    1.080 &   -1.490 \\
   23.00 &   5727.040 &    1.080 &   -2.605 \\
   23.00 &   5727.045 &    1.080 &   -0.914 \\
   23.00 &   5727.051 &    1.080 &   -1.417 \\
   23.00 &   5727.056 &    1.080 &   -2.400 \\
   23.00 &   5727.060 &    1.080 &   -1.043 \\
   23.00 &   5727.065 &    1.080 &   -1.411 \\
   23.00 &   5727.069 &    1.080 &   -2.303 \\
   23.00 &   5727.072 &    1.080 &   -1.189 \\
   23.00 &   5727.075 &    1.080 &   -1.458 \\
   23.00 &   5727.078 &    1.080 &   -2.303 \\
   23.00 &   5727.081 &    1.080 &   -1.359 \\
   23.00 &   5727.084 &    1.080 &   -1.563 \\
   23.00 &   5727.086 &    1.080 &   -2.458 \\
   23.00 &   5727.087 &    1.080 &   -1.563 \\
   23.00 &   5727.089 &    1.080 &   -1.759 \\
   23.00 &   5727.091 &    1.080 &   -1.826 \\
   23.00 &   5727.619 &    1.050 &   -1.456 \\
   23.00 &   5727.619 &    1.050 &   -1.867 \\
   23.00 &   5727.653 &    1.050 &   -1.753 \\
   23.00 &   5727.653 &    1.050 &   -2.072 \\
   23.00 &   5727.654 &    1.050 &   -1.867 \\
   23.00 &   5727.681 &    1.050 &   -1.753 \\
   23.00 &   5727.681 &    1.050 &   -1.878 \\
   23.00 &   5727.681 &    1.050 &   -9.850 \\
   23.00 &   5727.701 &    1.050 &   -2.054 \\
   23.00 &   5727.702 &    1.050 &   -1.878 \\
   23.00 &   6039.726 &    1.063 &   -0.650 \\
   23.00 &   6081.417 &    1.050 &   -1.814 \\
   23.00 &   6081.418 &    1.050 &   -1.638 \\
   23.00 &   6081.428 &    1.050 &   -1.638 \\
   23.00 &   6081.428 &    1.050 &   -9.610 \\
   23.00 &   6081.429 &    1.050 &   -1.513 \\
   23.00 &   6081.443 &    1.050 &   -1.513 \\
   23.00 &   6081.443 &    1.050 &   -1.832 \\
   23.00 &   6081.444 &    1.050 &   -1.627 \\
   23.00 &   6081.461 &    1.050 &   -1.627 \\
   23.00 &   6081.462 &    1.050 &   -1.216 \\
   23.00 &   6090.194 &    1.080 &   -0.700 \\
   23.00 &   6090.201 &    1.080 &   -0.841 \\
   23.00 &   6090.207 &    1.080 &   -1.005 \\
   23.00 &   6090.208 &    1.080 &   -1.540 \\
   23.00 &   6090.213 &    1.080 &   -1.203 \\
   23.00 &   6090.213 &    1.080 &   -1.344 \\
   23.00 &   6090.217 &    1.080 &   -1.290 \\
   23.00 &   6090.217 &    1.080 &   -1.458 \\
   23.00 &   6090.220 &    1.080 &   -2.655 \\
   23.00 &   6090.221 &    1.080 &   -1.312 \\
   23.00 &   6090.221 &    1.080 &   -1.846 \\
   23.00 &   6090.223 &    1.080 &   -1.403 \\
   23.00 &   6090.223 &    1.080 &   -2.244 \\
   23.00 &   6090.225 &    1.080 &   -1.591 \\
   23.00 &   6090.225 &    1.080 &   -2.022 \\
   23.00 &   6090.226 &    1.080 &   -1.897 \\
   23.00 &   6090.227 &    1.080 &   -1.846 \\
   23.00 &   6090.227 &    1.080 &   -1.876 \\
   23.00 &   6111.592 &    1.042 &   -1.701 \\
   23.00 &   6111.632 &    1.042 &   -1.224 \\
   23.00 &   6111.656 &    1.042 &   -1.224 \\
   23.00 &   6111.696 &    1.042 &   -1.370 \\
   23.00 &   6119.528 &    1.063 &   -0.360 \\
   23.00 &   6199.149 &    0.286 &   -2.133 \\
   23.00 &   6199.167 &    0.286 &   -2.238 \\
   23.00 &   6199.182 &    0.286 &   -2.354 \\
   23.00 &   6199.197 &    0.286 &   -2.483 \\
   23.00 &   6199.201 &    0.286 &   -3.141 \\
   23.00 &   6199.209 &    0.286 &   -2.629 \\
   23.00 &   6199.212 &    0.286 &   -2.930 \\
   23.00 &   6199.221 &    0.286 &   -2.799 \\
   23.00 &   6199.221 &    0.286 &   -2.857 \\
   23.00 &   6199.229 &    0.286 &   -2.851 \\
   23.00 &   6199.230 &    0.286 &   -3.003 \\
   23.00 &   6199.235 &    0.286 &   -2.898 \\
   23.00 &   6199.238 &    0.286 &   -3.266 \\
   23.00 &   6199.240 &    0.286 &   -3.003 \\
   23.00 &   6199.243 &    0.286 &   -3.199 \\
   23.00 &   6199.246 &    0.286 &   -4.443 \\
   23.00 &   6199.251 &    0.286 &   -4.045 \\
   23.00 &   6199.253 &    0.286 &   -3.840 \\
   23.00 &   6199.253 &    0.286 &   -3.898 \\
   23.00 &   6199.255 &    0.286 &   -3.743 \\
   23.00 &   6199.255 &    0.286 &   -3.743 \\
   23.00 &   6242.798 &    0.262 &   -2.054 \\
   23.00 &   6242.798 &    0.262 &   -2.521 \\
   23.00 &   6242.829 &    0.262 &   -2.375 \\
   23.00 &   6242.837 &    0.262 &   -2.375 \\
   23.00 &   6242.852 &    0.262 &   -2.396 \\
   23.00 &   6242.868 &    0.262 &   -2.852 \\
   23.00 &   6243.045 &    0.300 &   -2.712 \\
   23.00 &   6243.060 &    0.300 &   -2.497 \\
   23.00 &   6243.075 &    0.300 &   -2.420 \\
   23.00 &   6243.087 &    0.300 &   -1.649 \\
   23.00 &   6243.087 &    0.300 &   -2.409 \\
   23.00 &   6243.097 &    0.300 &   -1.785 \\
   23.00 &   6243.099 &    0.300 &   -2.452 \\
   23.00 &   6243.106 &    0.300 &   -1.933 \\
   23.00 &   6243.109 &    0.300 &   -2.555 \\
   23.00 &   6243.114 &    0.300 &   -2.092 \\
   23.00 &   6243.118 &    0.300 &   -2.776 \\
   23.00 &   6243.120 &    0.300 &   -2.261 \\
   23.00 &   6243.125 &    0.300 &   -2.428 \\
   23.00 &   6243.129 &    0.300 &   -2.566 \\
   23.00 &   6243.132 &    0.300 &   -2.580 \\
   23.00 &   6243.140 &    0.300 &   -2.712 \\
   23.00 &   6243.142 &    0.300 &   -2.776 \\
   23.00 &   6243.143 &    0.300 &   -2.497 \\
   23.00 &   6243.145 &    0.300 &   -2.555 \\
   23.00 &   6243.146 &    0.300 &   -2.420 \\
   23.00 &   6243.146 &    0.300 &   -2.452 \\
   23.00 &   6243.147 &    0.300 &   -2.409 \\
   23.00 &   6285.160 &    0.275 &   -1.540 \\
   24.00 &   4708.017 &    3.170 &    0.090 &    71.2 &    72.8 &    54.6 \\
   24.00 &   4767.860 &    3.560 &   -0.600 &    32.0 &    33.5 &    16.3 \\
   24.00 &   4789.340 &    2.540 &   -0.350 &    86.2 &    86.9 &    64.8 \\
   24.00 &   4801.047 &    3.120 &   -0.130 &    68.7 &    70.5 &    47.9 \\
   24.00 &   4936.335 &    3.110 &   -0.250 &    65.7 &    68.5 &    44.2 \\
   24.00 &   5214.140 &    3.370 &   -0.740 &    32.1 &    33.5 &    16.1 \\
   24.00 &   5238.964 &    2.710 &   -1.270 &    34.0 &    36.8 &    14.9 \\
   24.00 &   5247.566 &    0.960 &   -1.590 &   104.6 &   107.7 &    81.4 \\
   24.00 &   5272.007 &    3.450 &   -0.420 &    43.7 &    44.5 &    24.0 \\
   24.00 &   5287.200 &    3.440 &   -0.870 &    24.5 &    26.8 &    11.0 \\
   24.00 &   5628.621 &    3.420 &   -0.760 &    31.5 &    32.3 &    13.8 \\
   24.00 &   5783.080 &    3.320 &   -0.430 &    52.0 &    55.5 &    32.2 \\
   24.00 &   5783.870 &    3.320 &   -0.290 &    71.7 &    75.2 &    44.1 \\
   24.00 &   5787.930 &    3.322 &   -0.080 &    68.3 &    70.5 &    45.7 \\
   24.00 &   6330.100 &    0.941 &   -2.900 &    53.9 &    56.4 &    25.8 \\
   24.00 &   6882.477 &    3.438 &   -0.375 &    59.5 &    62.6 &    32.5 \\
   24.10 &   5237.328 &    4.070 &   -1.090 &    60.6 &    59.7 &    52.5 \\
   25.00 &   4709.690 &    2.886 &   -1.096 \\
   25.00 &   4709.698 &    2.886 &   -2.088 \\
   25.00 &   4709.698 &    2.886 &   -2.088 \\
   25.00 &   4709.705 &    2.886 &   -1.267 \\
   25.00 &   4709.711 &    2.886 &   -1.906 \\
   25.00 &   4709.711 &    2.886 &   -1.906 \\
   25.00 &   4709.717 &    2.886 &   -1.452 \\
   25.00 &   4709.722 &    2.886 &   -1.875 \\
   25.00 &   4709.723 &    2.886 &   -1.875 \\
   25.00 &   4709.728 &    2.886 &   -1.644 \\
   25.00 &   4709.731 &    2.886 &   -1.940 \\
   25.00 &   4709.731 &    2.886 &   -1.940 \\
   25.00 &   4709.735 &    2.886 &   -1.819 \\
   25.00 &   4709.737 &    2.886 &   -2.138 \\
   25.00 &   4709.737 &    2.886 &   -2.138 \\
   25.00 &   4709.740 &    2.886 &   -1.883 \\
   25.00 &   4739.068 &    2.939 &   -1.632 \\
   25.00 &   4739.069 &    2.939 &   -1.155 \\
   25.00 &   4739.087 &    2.939 &   -1.530 \\
   25.00 &   4739.088 &    2.939 &   -1.704 \\
   25.00 &   4739.089 &    2.939 &   -1.632 \\
   25.00 &   4739.101 &    2.939 &   -1.662 \\
   25.00 &   4739.102 &    2.939 &   -3.240 \\
   25.00 &   4739.103 &    2.939 &   -1.530 \\
   25.00 &   4739.111 &    2.939 &   -2.030 \\
   25.00 &   4739.112 &    2.939 &   -1.662 \\
   25.00 &   5004.892 &    2.918 &   -1.630 \\
   26.00 &   4745.800 &    3.650 &   -1.270 &    97.2 &    99.1 &    77.3 \\
   26.00 &   4749.950 &    4.560 &   -1.240 &    55.4 &    56.6 &    35.9 \\
   26.00 &   4799.410 &    3.640 &   -2.130 &    52.8 &    53.5 &    33.4 \\
   26.00 &   4808.150 &    3.250 &   -2.690 &    43.2 &    44.3 &    26.0 \\
   26.00 &   4973.090 &    3.960 &   -0.770 &   103.6 &   108.2 &    82.6 \\
   26.00 &   5044.211 &    2.850 &   -2.060 &    93.7 &    95.6 &    73.0 \\
   26.00 &   5054.642 &    3.640 &   -1.920 &    61.8 &    62.2 &    40.3 \\
   26.00 &   5067.140 &    4.220 &   -0.860 &    93.2 &    94.7 &    67.8 \\
   26.00 &   5127.679 &    0.050 &   -6.120 &    39.4 &    42.0 &    16.9 \\
   26.00 &   5187.910 &    4.140 &   -1.260 &    80.5 &    82.2 &    58.3 \\
   26.00 &   5225.525 &    0.110 &   -4.790 &    95.5 &    98.0 &    71.8 \\
   26.00 &   5250.208 &    0.120 &   -4.940 &    85.8 &    88.1 &    64.6 \\
   26.00 &   5253.460 &    3.280 &   -1.570 &   101.2 &   101.8 &    79.2 \\
   26.00 &   5409.130 &    4.370 &   -1.060 &    77.0 &    76.8 &    57.7 \\
   26.00 &   5466.987 &    3.570 &   -2.230 &    54.7 &    54.9 &    32.7 \\
   26.00 &   5577.020 &    5.030 &   -1.460 &    20.8 &    22.1 &    10.4 \\
   26.00 &   5636.696 &    3.640 &   -2.560 &    35.0 &    35.7 &    18.8 \\
   26.00 &   5638.262 &    4.220 &   -0.770 &   101.6 &   101.1 &    76.1 \\
   26.00 &   5649.987 &    5.100 &   -0.800 &    54.3 &    54.6 &    34.8 \\
   26.00 &   5651.469 &    4.470 &   -1.750 &    34.1 &    34.0 &    18.2 \\
   26.00 &   5661.348 &    4.280 &   -1.760 &    41.5 &    42.7 &    22.5 \\
   26.00 &   5679.023 &    4.650 &   -0.750 &    76.6 &    78.9 &    57.9 \\
   26.00 &   5696.089 &    4.550 &   -1.780 &    26.7 &    27.1 &    13.0 \\
   26.00 &   5705.464 &    4.300 &   -1.360 &    56.4 &    55.8 &    36.8 \\
   26.00 &   5731.760 &    4.260 &   -1.200 &    76.8 &    78.4 &    57.4 \\
   26.00 &   5778.453 &    2.590 &   -3.440 &    43.2 &    44.5 &    21.8 \\
   26.00 &   5784.660 &    3.400 &   -2.530 &    46.2 &    48.6 &    25.5 \\
   26.00 &   5793.914 &    4.220 &   -1.620 &    51.3 &    53.4 &    32.1 \\
   26.00 &   5806.730 &    4.610 &   -0.950 &    76.8 &    79.1 &    52.9 \\
   26.00 &   5852.220 &    4.550 &   -1.230 &    62.3 &    62.8 &    39.5 \\
   26.00 &   5855.076 &    4.610 &   -1.480 &    39.3 &    40.3 &    22.5 \\
   26.00 &   5856.090 &    4.290 &   -1.460 &    51.8 &    52.5 &    32.9 \\
   26.00 &   5927.789 &    4.650 &   -1.040 &    60.1 &    59.9 &    41.6 \\
   26.00 &   5934.655 &    3.930 &   -1.070 &    99.9 &   101.2 &    75.9 \\
   26.00 &   6056.005 &    4.730 &   -0.400 &    92.2 &    93.9 &    71.4 \\
   26.00 &   6079.009 &    4.650 &   -1.020 &    64.7 &    63.9 &    44.7 \\
   26.00 &   6082.711 &    2.220 &   -3.570 &    55.0 &    57.6 &    33.8 \\
   26.00 &   6093.644 &    4.610 &   -1.300 &    49.1 &    48.5 &    29.6 \\
   26.00 &   6127.910 &    4.140 &   -1.400 &    69.5 &    68.3 &    49.6 \\
   26.00 &   6151.618 &    2.180 &   -3.280 &    70.3 &    71.5 &    49.0 \\
   26.00 &   6157.728 &    4.080 &   -1.220 &    79.9 &    81.3 &    60.8 \\
   26.00 &   6165.360 &    4.140 &   -1.460 &    61.7 &    63.7 &    43.1 \\
   26.00 &   6213.430 &    2.220 &   -2.520 &   102.8 &   106.8 &    81.3 \\
   26.00 &   6219.281 &    2.200 &   -2.430 &   116.5 &   118.3 &    88.7 \\
   26.00 &   6226.736 &    3.880 &   -2.100 &    49.4 &    49.9 &    28.5 \\
   26.00 &   6252.555 &    2.400 &   -1.690 &   153.3 &   157.8 &   118.9 \\
   26.00 &   6270.225 &    2.860 &   -2.540 &    73.7 &    75.2 &    51.2 \\
   26.00 &   6271.279 &    3.330 &   -2.700 &    41.3 &    41.8 &    22.9 \\
   26.00 &   6335.330 &    2.200 &   -2.260 &   121.7 &   123.9 &    95.9 \\
   26.00 &   6392.539 &    2.280 &   -4.030 &    35.2 &    36.4 &    15.8 \\
   26.00 &   6481.870 &    2.280 &   -2.980 &    85.7 &    87.0 &    63.6 \\
   26.00 &   6518.370 &    2.830 &   -2.450 &    76.7 &    78.6 &    56.2 \\
   26.00 &   6593.871 &    2.430 &   -2.390 &   109.4 &   110.2 &    82.7 \\
   26.00 &   6597.561 &    4.800 &   -0.970 &    62.3 &    63.1 &    43.2 \\
   26.00 &   6625.022 &    1.010 &   -5.340 &    34.7 &    36.6 &    14.8 \\
   26.00 &   6699.142 &    4.593 &   -2.101 &    18.4 &    18.8 &     8.1 \\
   26.00 &   6703.567 &    2.760 &   -3.020 &    60.7 &    61.7 &    36.8 \\
   26.00 &   6705.102 &    4.610 &   -0.980 &    69.2 &    70.3 &    46.0 \\
   26.00 &   6713.745 &    4.800 &   -1.400 &    37.7 &    37.9 &    20.5 \\
   26.00 &   6725.357 &    4.100 &   -2.190 &    33.2 &    34.6 &    16.5 \\
   26.00 &   6726.667 &    4.610 &   -1.030 &    66.0 &    66.4 &    46.5 \\
   26.00 &   6733.151 &    4.640 &   -1.470 &    45.9 &    47.2 &    26.1 \\
   26.00 &   6750.152 &    2.420 &   -2.620 &    98.2 &   100.9 &    73.9 \\
   26.00 &   6806.845 &    2.730 &   -3.110 &    58.1 &    60.1 &    34.2 \\
   26.00 &   6810.263 &    4.610 &   -0.990 &    70.3 &    70.5 &    49.5 \\
   26.00 &   6828.590 &    4.640 &   -0.820 &    77.2 &    77.8 &    54.3 \\
   26.00 &   6837.006 &    4.590 &   -1.690 &    33.3 &    34.1 &    17.9 \\
   26.00 &   6842.690 &    4.640 &   -1.220 &    58.3 &    58.0 &    36.7 \\
   26.00 &   6843.656 &    4.550 &   -0.830 &    83.8 &    83.8 &    60.4 \\
   26.00 &   6858.150 &    4.610 &   -0.940 &    69.9 &    70.9 &    50.9 \\
   26.00 &   6999.880 &    4.100 &   -1.460 &    75.1 &    75.5 &    54.0 \\
   26.00 &   7132.990 &    4.080 &   -1.650 &    63.0 &    63.1 &    41.8 \\
   26.00 &   7401.685 &    4.186 &   -1.500 &    60.0 &    59.9 &    40.5 \\
   26.00 &   7418.670 &    4.140 &   -1.380 &    71.2 &    71.5 &    47.5 \\
   26.10 &   4620.510 &    2.830 &   -3.210 &    60.0 &    57.9 &    52.4 \\
   26.10 &   4993.340 &    2.810 &   -3.730 &    44.3 &    43.4 &    38.3 \\
   26.10 &   5414.073 &    3.220 &   -3.580 &    34.3 &    32.8 &    26.6 \\
   26.10 &   6084.111 &    3.200 &   -3.830 &    25.9 &    24.8 &    20.1 \\
   26.10 &   6416.919 &    3.890 &   -2.750 &    47.2 &    45.4 &    38.5 \\
   26.10 &   6432.680 &    2.890 &   -3.570 &    46.7 &    45.7 &    39.5 \\
   26.10 &   6456.383 &    3.900 &   -2.050 &    67.4 &    63.6 &    61.7 \\
   26.10 &   7711.721 &    3.903 &   -2.500 &    53.1 &    50.2 &    44.5 \\
   27.00 &   4792.846 &    3.250 &   -0.070 \\
   27.00 &   4813.467 &    3.213 &    0.050 \\
   27.00 &   5212.691 &    3.512 &   -0.110 \\
   27.00 &   5247.911 &    1.784 &   -2.070 \\
   27.00 &   5647.234 &    2.278 &   -1.560 \\
   27.00 &   6093.143 &    1.739 &   -2.440 \\
   27.00 &   6454.990 &    3.629 &   -0.250 \\
   28.00 &   4831.180 &    3.610 &   -0.320 &    91.9 &    92.4 &    73.2 \\
   28.00 &   4866.270 &    3.540 &   -0.210 &   102.4 &   101.3 &    77.1 \\
   28.00 &   4913.980 &    3.740 &   -0.660 &    75.2 &    75.2 &    55.7 \\
   28.00 &   4946.040 &    3.800 &   -1.220 &    51.2 &    53.0 &    28.0 \\
   28.00 &   4952.290 &    3.610 &   -1.260 &    54.5 &    55.7 &    32.4 \\
   28.00 &   4953.208 &    3.740 &   -0.660 &    75.6 &    76.2 &    56.0 \\
   28.00 &   4976.135 &    3.610 &   -1.250 &    51.3 &    49.7 &    29.0 \\
   28.00 &   5010.938 &    3.640 &   -0.870 &    65.0 &    65.6 &    48.4 \\
   28.00 &   5082.350 &    3.660 &   -0.590 &    99.3 &    96.8 &    67.0 \\
   28.00 &   5084.110 &    3.680 &   -0.060 &   108.3 &   109.8 &    88.1 \\
   28.00 &   5094.420 &    3.830 &   -1.070 &    47.7 &    47.3 &    28.7 \\
   28.00 &   5157.980 &    3.610 &   -1.510 &    34.4 &    35.2 &    16.9 \\
   28.00 &   5578.729 &    1.680 &   -2.570 &    80.5 &    83.1 &    57.4 \\
   28.00 &   5589.358 &    3.900 &   -1.140 &    45.2 &    45.7 &    28.1 \\
   28.00 &   5593.746 &    3.900 &   -0.780 &    64.7 &    66.0 &    43.8 \\
   28.00 &   5625.320 &    4.090 &   -0.730 &    58.4 &    56.6 &    37.1 \\
   28.00 &   5628.350 &    4.090 &   -1.320 &    32.1 &    30.8 &    14.3 \\
   28.00 &   5638.750 &    3.900 &   -1.700 &    22.7 &    22.2 &     9.6 \\
   28.00 &   5641.880 &    4.110 &   -1.020 &    44.8 &    44.9 &    23.2 \\
   28.00 &   5643.078 &    4.160 &   -1.250 &    29.9 &    30.0 &    14.6 \\
   28.00 &   5694.990 &    4.090 &   -0.630 &    62.0 &    62.7 &    42.7 \\
   28.00 &   5748.360 &    1.680 &   -3.240 &    51.1 &    52.2 &    27.5 \\
   28.00 &   5754.670 &    1.930 &   -1.850 &    99.5 &   101.9 &    78.7 \\
   28.00 &   5805.217 &    4.170 &   -0.640 &    61.6 &    63.6 &    40.4 \\
   28.00 &   5847.010 &    1.676 &   -3.410 &    46.2 &    48.0 &    22.6 \\
   28.00 &   6086.282 &    4.270 &   -0.510 &    67.1 &    68.6 &    43.4 \\
   28.00 &   6111.080 &    4.088 &   -0.810 &    58.5 &    58.4 &    33.3 \\
   28.00 &   6119.760 &    4.270 &   -1.316 &    23.5 &    24.2 &    10.6 \\
   28.00 &   6128.984 &    1.677 &   -3.360 &    50.1 &    49.0 &    25.0 \\
   28.00 &   6130.135 &    4.270 &   -0.960 &    41.4 &    40.5 &    21.9 \\
   28.00 &   6175.370 &    4.089 &   -0.550 &    69.1 &    69.2 &    49.0 \\
   28.00 &   6176.811 &    4.090 &   -0.260 &    87.5 &    87.9 &    62.9 \\
   28.00 &   6177.242 &    1.830 &   -3.510 &    34.2 &    33.7 &    13.5 \\
   28.00 &   6186.717 &    4.110 &   -0.960 &    52.0 &    52.6 &    30.6 \\
   28.00 &   6204.604 &    4.090 &   -1.140 &    42.2 &    43.1 &    21.7 \\
   28.00 &   6223.971 &    4.105 &   -1.466 &    47.8 &    48.8 &    27.0 \\
   28.00 &   6230.100 &    4.110 &   -1.132 &    40.3 &    42.4 &    19.3 \\
   28.00 &   6322.169 &    4.154 &   -1.210 &    36.6 &    36.6 &    18.5 \\
   28.00 &   6360.810 &    4.170 &   -1.150 &    37.0 &    35.1 &    16.6 \\
   28.00 &   6378.233 &    4.154 &   -1.386 &    55.0 &    55.4 &    31.0 \\
   28.00 &   6598.611 &    4.236 &   -0.910 &    43.2 &    43.9 &    24.8 \\
   28.00 &   6635.130 &    4.420 &   -0.720 &    44.8 &    44.3 &    23.3 \\
   28.00 &   6643.630 &    1.680 &   -2.000 &   126.8 &   126.5 &    92.1 \\
   28.00 &   6767.772 &    1.830 &   -2.170 &   103.4 &   106.2 &    78.4 \\
   28.00 &   6772.315 &    3.660 &   -0.990 &    72.6 &    73.9 &    49.2 \\
   28.00 &   6842.043 &    3.658 &   -1.500 &    45.0 &    40.8 &    24.2 \\
   28.00 &   7715.591 &    3.700 &   -1.010 &    76.3 &    77.0 &    50.0 \\
   28.00 &   7727.624 &    3.680 &   -0.400 &   117.5 &   118.2 &    91.0 \\
   28.00 &   7748.890 &    3.700 &   -0.380 &   115.6 &   116.4 &    85.0 \\
   28.00 &   7797.586 &    3.890 &   -0.340 &   104.0 &   107.2 &    74.0 \\
   29.00 &   5218.197 &    3.814 &    0.480 \\
   29.00 &   7933.096 &    3.783 &   -0.877 \\
   29.00 &   7933.098 &    3.783 &   -0.877 \\
   29.00 &   7933.119 &    3.783 &   -0.877 \\
   29.00 &   7933.119 &    3.783 &   -0.877 \\
   29.00 &   7933.134 &    3.783 &   -1.576 \\
   29.00 &   7933.135 &    3.783 &   -1.576 \\
   29.00 &   7933.155 &    3.783 &   -0.877 \\
   29.00 &   7933.157 &    3.783 &   -0.877 \\
   38.00 &   4607.338 &    0.000 &    0.283 &    68.5 &    69.2 &    46.9 \\
   39.10 &   4854.867 &    0.992 &   -0.380 &    55.7 &    55.1 &    47.7 \\
   39.10 &   4883.685 &    1.084 &    0.070 &    62.3 &    62.7 &    56.8 \\
   39.10 &   4900.110 &    1.033 &   -0.090 &    61.7 &    60.1 &    54.1 \\
   39.10 &   5087.420 &    1.084 &   -0.170 &    58.3 &    57.9 &    47.0 \\
   39.10 &   5200.413 &    0.992 &   -0.570 &    47.8 &    47.2 &    37.9 \\
   56.00 &   5853.686 &    0.604 &   -2.066 \\
   56.00 &   5853.687 &    0.604 &   -2.066 \\
   56.00 &   5853.687 &    0.604 &   -2.009 \\
   56.00 &   5853.688 &    0.604 &   -2.009 \\
   56.00 &   5853.689 &    0.604 &   -2.215 \\
   56.00 &   5853.689 &    0.604 &   -2.215 \\
   56.00 &   5853.690 &    0.604 &   -1.010 \\
   56.00 &   5853.690 &    0.604 &   -1.466 \\
   56.00 &   5853.690 &    0.604 &   -1.914 \\
   56.00 &   5853.690 &    0.604 &   -2.620 \\
   56.00 &   5853.690 &    0.604 &   -1.010 \\
   56.00 &   5853.690 &    0.604 &   -1.466 \\
   56.00 &   5853.690 &    0.604 &   -1.914 \\
   56.00 &   5853.690 &    0.604 &   -2.620 \\
   56.00 &   5853.690 &    0.604 &   -1.010 \\
   56.00 &   5853.691 &    0.604 &   -2.215 \\
   56.00 &   5853.692 &    0.604 &   -2.215 \\
   56.00 &   5853.693 &    0.604 &   -2.009 \\
   56.00 &   5853.693 &    0.604 &   -2.009 \\
   56.00 &   5853.694 &    0.604 &   -2.066 \\
   56.00 &   5853.694 &    0.604 &   -2.066 \\
   56.00 &   6141.725 &    0.704 &   -2.456 \\
   56.00 &   6141.725 &    0.704 &   -2.456 \\
   56.00 &   6141.727 &    0.704 &   -1.311 \\
   56.00 &   6141.727 &    0.704 &   -1.311 \\
   56.00 &   6141.728 &    0.704 &   -2.284 \\
   56.00 &   6141.728 &    0.704 &   -2.284 \\
   56.00 &   6141.729 &    0.704 &   -0.503 \\
   56.00 &   6141.729 &    0.704 &   -1.214 \\
   56.00 &   6141.729 &    0.704 &   -0.503 \\
   56.00 &   6141.729 &    0.704 &   -1.214 \\
   56.00 &   6141.730 &    0.704 &   -0.077 \\
   56.00 &   6141.730 &    0.704 &   -0.077 \\
   56.00 &   6141.730 &    0.704 &   -0.077 \\
   56.00 &   6141.731 &    0.704 &   -0.709 \\
   56.00 &   6141.731 &    0.704 &   -1.327 \\
   56.00 &   6141.731 &    0.704 &   -0.709 \\
   56.00 &   6141.731 &    0.704 &   -1.327 \\
   56.00 &   6141.732 &    0.704 &   -0.959 \\
   56.00 &   6141.732 &    0.704 &   -1.281 \\
   56.00 &   6141.732 &    0.704 &   -0.959 \\
   56.00 &   6141.733 &    0.704 &   -1.281 \\
   56.00 &   6496.898 &    0.604 &   -1.886 \\
   56.00 &   6496.899 &    0.604 &   -1.886 \\
   56.00 &   6496.901 &    0.604 &   -1.186 \\
   56.00 &   6496.902 &    0.604 &   -1.186 \\
   56.00 &   6496.906 &    0.604 &   -0.739 \\
   56.00 &   6496.906 &    0.604 &   -0.739 \\
   56.00 &   6496.910 &    0.604 &   -0.380 \\
   56.00 &   6496.910 &    0.604 &   -0.380 \\
   56.00 &   6496.910 &    0.604 &   -0.380 \\
   56.00 &   6496.916 &    0.604 &   -1.583 \\
   56.00 &   6496.916 &    0.604 &   -1.583 \\
   56.00 &   6496.917 &    0.604 &   -1.186 \\
   56.00 &   6496.918 &    0.604 &   -1.186 \\
   56.00 &   6496.920 &    0.604 &   -1.186 \\
   56.00 &   6496.922 &    0.604 &   -1.186 \\
\label{linelist}
\end{longtable}
}

\end{document}